\documentclass[pra,amsmath,superscriptaddress,twocolumn,showpacs]{revtex4-1}
\usepackage{bm}
\usepackage{amssymb}
\usepackage{colordvi}
\usepackage{graphicx}
\usepackage{color}
\usepackage{hyperref}

\newcommand{\be}{\begin{equation}}
\newcommand{\ee}{\end{equation}}
\newcommand{\bea}{\begin{eqnarray}}
\newcommand{\eea}{\end{eqnarray}}

\newcommand{\vep}{\varepsilon}
\newcommand{\ave}[1]{\langle #1\rangle}
\newcommand{\ome}{\omega}

\def\Im {\mbox{Im}}

\def\nn{\nonumber}

\begin{document}

\title{Near-field three-terminal thermoelectric heat engine}

\author{Jian-Hua Jiang}
\affiliation{College of Physics, Optoelectronics and Energy, \&
  Collaborative Innovation Center of Suzhou Nano Science and
  Technology, Soochow University, 1 Shizi Street, Suzhou 215006,
  China}
\author{Yoseph Imry}
\affiliation{Department of Condensed Matter Physics, Weizmann Institute of
  Science, Rehovot 76100, Israel}

\date{\today}

\begin{abstract}
We propose a near-field inelastic thermoelectric heat engine where quantum-dots are used to effectively rectify the charge flow
of photo-carriers. The device converts {near-field heat radiation} into useful electrical
power. Heat absorption and inelastic transport can be enhanced by introducing two continuous spectra {separated
by an energy gap}. The thermoelectric transport properties of the heat engine are studied in the linear-response 
regime. {Using a small band-gap semiconductor as the absorption material, we show that the device achieves
very large thermopower and thermoelectric figure-of-merit, as well as considerable power-factor. By analyzing thermal-photo-carrier
generation and conduction, we reveal that the Seebeck coefficient and the figure of merit have oscillatory dependence on
the thickness of the vacuum gap. Meanwhile, the power-factor, the charge and thermal 
conductivity are significantly improved by near-field radiation. Conditions and guiding principles for powerful and efficient
thermoelectric heat engines are discussed in details.}
\end{abstract}


\maketitle

\section{Introduction}
Thermoelectric phenomena are at the heart of {transport study of material properties which reveals important
information of microscopic quasiparticle processes} \cite{te,joe}. Being able to generate 
electrical power using a temperature gradient or cool an object using electrical power without mechanical motion or 
chemical processes, thermoelectricity {has been one of the main branches in renewable energy research for
decades \cite{hicks,MS,dressel,zhao}. The main challenges here are to improve the energy efficiency and power density, 
using low-cost and environment-friendly materials and designs.} 

Mahan and Sofo pointed out that the thermoelectric figure of merit can be written as \cite{MS}
\be
ZT = \frac{\sigma S^2T}{\kappa_e+\kappa_p}= \frac{\ave{E-\mu}^2}{{\rm Var}(E-\mu)+\Lambda^2}, \ \ \Lambda=e^2\kappa_p T/\sigma , \label{ms}
\ee
where $\sigma$ is the electrical conductivity, $S$ is the Seebeck coefficient, $T$ is the equilibrium temperature,
$\kappa_e$ is the electronic heat conductivity, and $\kappa_p$ is the heat conductivity due to phonons.
The last equality rewrites the figure of merit in microscopic quantities. The key is to assume an energy-dependent
conductivity $\sigma(E)$, and the average here is defined as \cite{MS}
\be
\ave{...} = \frac{\int dE \sigma(E) (-\partial_E f(E)) ...}{\int dE \sigma(E) (-\partial_E f(E))} ,
\ee
where $f(E)=[\exp(\frac{E-\mu}{k_BT})+1]^{-1}$ is the Fermi distribution function with $\mu$ being
the equilibrium chemical potential. {The symbol ``${\rm Var}$" in Eq.~(\ref{ms}) denotes the variance, 
${\rm Var}(E-\mu)=\ave{(E-\mu)^2}-\ave{E-\mu}^2$.}
Eq.~(\ref{ms}) shows that without phonon heat conductivity,
the thermoelectric figure of merit is proportional to the square of the average of electronic heat $E-\mu$ over its
variance. Based on these observations, Mahan and Sofo \cite{MS} proposed to achieve high figure of merit by 
using {materials with narrow band-width conduction band (such as $f$-electron bands)} 
to reduce the variance of electronic heat $E-\mu$, while keeping a decent average value.
{However, in reality, narrow-band semiconductors have very small electrical conductivity, hence the
factor $\Lambda$ is significantly increased, which instead results in much reduced figure 
of merit \cite{zhou,low-diss}.} Meanwhile, the power obtained is also small if the electrical conductivity is small \cite{low-diss,kedem,amnon,cb}. 
These factors reveal that the intrinsic entanglement between heat and electrical conductivity, as well as the parasitic phonon heat
conduction, impede the improvement of thermoelectric energy efficiency.

One approach to improve thermoelectric performance is to use inelastic transport processes,
where electrical and heat currents are carried by different quasiparticles and separated spatially. By inelastic
transport processes we mean that the transported electrons exchange energy with another degree of
freedom (characterized by some collective excitations, e.g., phonons) during the transport processes. Such 
effects also take place in conventional thermoelectric materials, but play only a minor role since the 
elastic transport processes are always dominant. When, e.g., elastic transport processes
are suppressed by an energy barrier, the inelastic processes become dominant. A typical set-up for inelastic
thermoelectricity is a three-terminal device, where the third terminal supplies the collective excitations (which
can often be described by some kind of bosonic quasi-particles) from a thermal
reservoir (e.g., see Fig.~1). It has been shown that the inelastic thermoelectric 
figure of merit is given by \cite{prb1,pn,rev}
\be
Z_{in}T = \frac{\ave{\hbar\ome}^2}{{\rm Var}(\hbar\ome)+\Lambda_{in}} \label{3tzt}
\ee
where $\hbar\ome$ denotes the energy of bosons that assist the inelastic transport and $\Lambda_{in}$ measures
the useless parasitic heat conduction. The above average is defined as
\be
\ave{\hbar\ome} = \frac{\int \hbar d\ome G_{in}(\hbar\ome) \hbar\ome}{\int \hbar d\ome G_{in}(\hbar\ome)} .
\ee
where the boson-energy dependent conductance $G_{in}(\hbar\ome)$ already includes electron and boson distribution
functions. To obtain high figure of merit through inelastic thermoelectricity, a narrow 
band-width of the bosons is needed, which, however, does not conflict with a high electrical conductivity. 
The above analysis thus opens the possibility of realizing high performance thermoelectric materials using inelastic 
thermoelectric transport. In the past years, studies have revealed the advantages of inelastic thermoelectric materials and devices \cite{cam,edwards,3t0,3t1,3tp1,3tp2,3t2,3t3,3t4,3t5,3t6,3t7,3tjap,3t8,3t9,3t10,3t11,3t12,3t13,3tpre,3tora,3t14,3t15,3tp3,3t16,3t17,3t18,3t19,3t20,3t21,3tpra,3t22,3t23,exp1,exp2,exp3}.

In this work, we propose a type of near-field inelastic three-terminal (NFI3T) quantum-dots (QDs) heat engine. The structure and working 
principle of the device is schematically illustrated in Fig.~1. The near field enhanced heat flux, in the form of infrared photons, is injected into
the device, absorbed, and converted into useful electrical power. The underlying mechanisms are the photo-carrier generation in
the central region and hot-carrier extraction through two QDs tunneling layers at the left and right sides. This device 
takes the following advantages of near-field radiations: First, the near-field radiation can strongly enhance heat transfer across the
vacuum gap \cite{nf-1} (for recent experiments,  see Refs.~\onlinecite{nf0,nf1,nf2,nf3,nf4,nf5,nf6}) and thus leads to significant 
heat flux injection. The injected infrared photons with energy greater than the energy-gap in the absorption material, $E_g$, 
lead to generation of photo-carriers and electrical power output. The larger the injected heat flux is, the larger is the output power. 
Second, unlike phonon-assisted interband transitions, photon-assisted interband transition is not limited by the small phonon 
frequency and can work for larger band-gaps {due to the continuous photon spectrum}.

In order to elaborate on the design principles, thermoelectric transport properties, and merits of the NFI3T heat engine, this paper is 
organized as follows: In Sec.~II we introduce the working principles of the NFI3T heat engine. In Sec.~III we 
obtain the transport equations from the linear-response theory. In Sec.~IV we study the transport properties and thermoelectric
energy conversion from microscopic theory. In Sec.~V we discuss the performance of the NFI3T heat engine. {We conclude and give 
an outlook in Sec.~VI. Our study here is mainly focused on the linear-response regime where a direct comparison with conventional
thermoelectricity is available.}

\section{Working principles of the near-field three-terminal heat engine}

\begin{figure}[htb]
   \includegraphics[width=6cm]{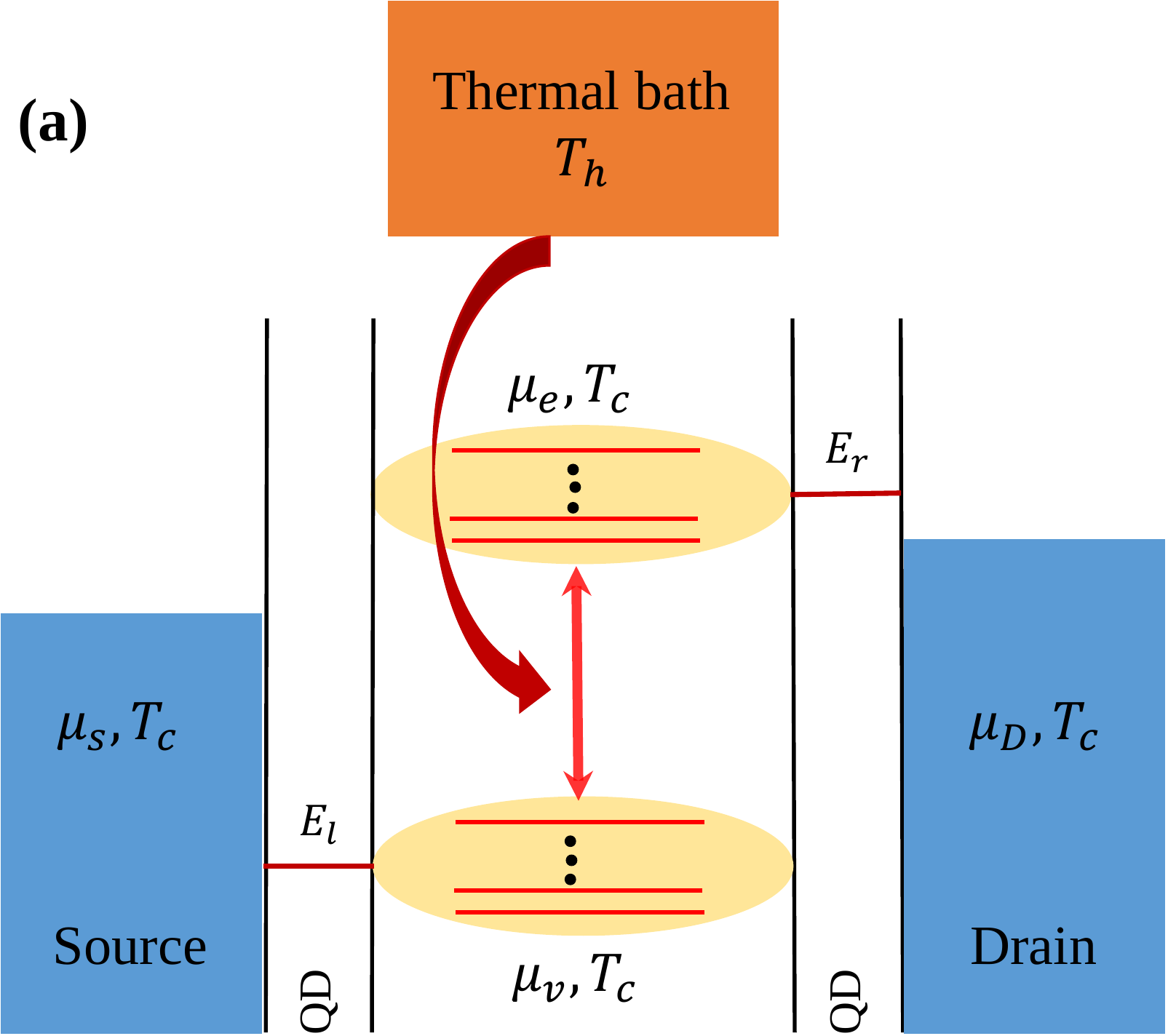}\\
   \vskip 0.6cm
  \includegraphics[width=6cm]{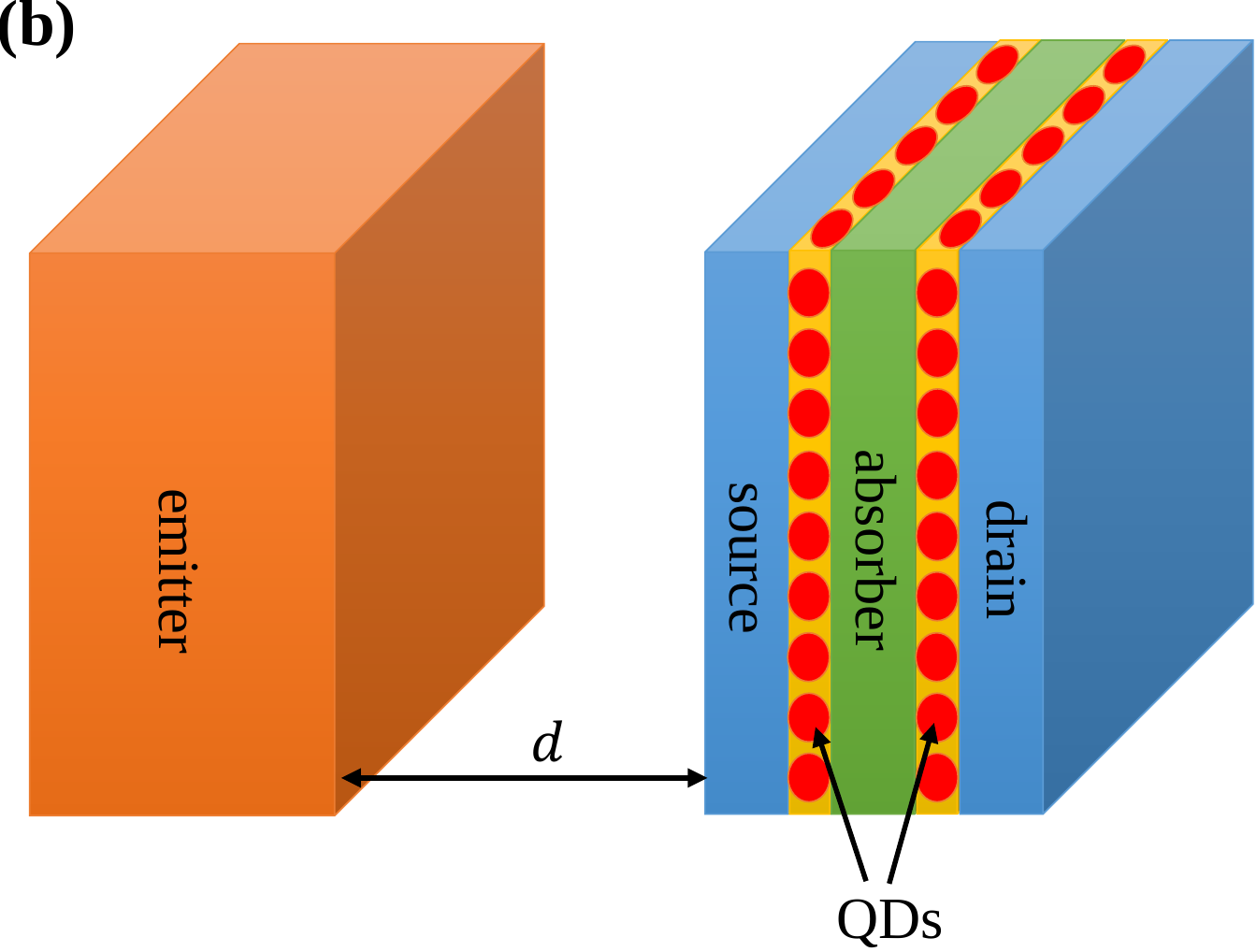}
  \caption{(Color online) {(a) Schematic of near-field three-terminal thermoelectric heat engine. A hot thermal reservoir of temperature 
  $T_h$ injects heat flux into the device through near-field heat radiation. The device is held at a lower temperature $T_c$. The absorption of the heat 
   radiation is realized by photon-assisted transitions between the two continua. As a result, the upper and lower continua have different chemical 
   potentials, $\mu_e$ and $\mu_v$, respectively. The source and drain have different electrochemical potentials, denoted as $\mu_S$ and $\mu_D$, 
   separately. Extracting the photo-thermal-carriers through the two QDs layers can lead to useful electrical power output. The typical energy of 
   QDs in the left (right) layer is $E_{\ell}$ ($E_{r}$). The energy levels of the left (right) QDs layer match the lower (upper) continuum in the absorption
   material to induce resonant tunneling. (b) A possible set-up for the three-terminal near-field heat engine. The emitter is a heat source of temperature $T_h$ which is separated from the device by a vacuum gap of thickness $d$. The device is held at a lower temperature $T_c$ which consists of the source, drain, and absorber layers. These three parts are divided by two layers of quantum dots arrays.}}
\label{fig1}
\end{figure}

The system we consider is schematically illustrated in Fig. 1(a). At the center of the engine,
there are two continuous spectra of electrons which can be realized by the valence and conduction 
bands in semiconductors, or by the mini-bands in superlattices, or anything alike. The equilibrium chemical
potential is set in between the two continua. By absorbing infrared photons from the thermal bath, 
electrons in the lower continuum jump into the upper continuum. The excess electrons occupying the
upper continuum will flow into the drain terminal, while the photon-generated holes in the lower continuum will
be refilled by electrons from the source. Since the lower continuum is strongly 
coupled with the source, whereas the upper continuum is strongly coupled with the drain, the hot photo-carrier diffusion
leads to a directional electrical current. Such selective strong coupling between the lower (upper) continuum and the 
source (drain) can be realized using resonant-tunneling through QDs. This is because the lowest energy level 
and the level-spacing in QDs can be controlled by the material and the quantum confinement effect (both conduction
and valence band energy levels are usable). 
To improve the harvesting of heat from the thermal terminal (a reservoir 
with high temperature $T_h$), we utilize the near-field thermal radiation across the vacuum gap between
the thermal emitter (i.e., the thermal terminal) and the device. It was known a long time ago \cite{nf-1} that when two objects are spaced 
closely, the evanescent electromagnetic waves can dramatically enhance the heat transfer between them.
In recent decades, near-field heat transfer across nanoscale vacuum gaps was realized and measured in
different experimental systems \cite{nf-1,nf0,nf1,nf2,nf3,nf4,nf5,nf6}. Spectral tailoring of near-field radiation is proposed by 
using surface electromagnetic waves \cite{nftpv0,nftpv1,nftpv2,nftpv3,nf-pc1,nf-pc2,nf-pc3,nf-pc4,nf-pc5}, meanwhile the 
light-matter interaction can also be strongly enhanced through these surface waves \cite{nftpv0,nftpv1,nftpv2,nftpv3,nf-pc1,nf-pc2,nf-pc3,nf-pc4,nf-pc5,nf-pc6}.
In the NFI3T heat engine, the use of near field heat transfer has several advantages. One of them is to enhance the
heat injection flux so that the output power of the heat engine can be considerably increased. {On the other hand, since near-field heat 
transfer is determined by the optical properties of the emitters and receivers, the near-field heat transfer becomes more controllable 
than heat diffusion in all-solid devices. For instance, the parasitic heat conduction can be reduced by manipulating the infrared thermal 
radiation using optical materials or meta-materials.} The resonant tunneling between the lower (upper) continuum 
and the source (drain) can be made efficient, if there are many QDs with energies matching the lower (upper) 
continuum (e.g., using QDs arrays). We remark that the NFI3T heat engine proposed here is different from the near-field thermophotovoltaic
devices in several aspects. First, thermophotovoltaic devices work in the temperature range of $1000$~K$<T_h<2000$~K,
while our device is designed for applications with $T_h<800$~K which fits most of {the industrial and} daily-life
renewable energy demands {for waste heat harvesting}. Besides, the transport mechanism here is different from the thermophotovoltaic
devices. Instead of the $p$-$n$ junction-type transport in thermophotovoltaic devices, here we have the resonant tunneling through
the QDs as one of the main transport mechanisms.

From electronic aspects, the proposed device has the following merit, compared to the inelastic thermoelectric devices in
the existing studies \cite{rev}. The use of the two continua can significantly
increase the photon absorption rate, compared to previous devices where direct photon-assited (or phonon-assisted) 
hopping between QDs are considered \cite{3tp1,3tp2,3tp3,prb1} (i.e., without the continua in the center). This 
improvement comes mainly from two aspects: First, by introducing the two continua in the center, the electronic densities of states 
are considerably increased. Second, we notice that the direct transition matrix element between two QDs assisted by photons is
suppressed by an exponential factor $\sim \exp(-2d/\xi)$ where $d$ is the distance between the QDs and $\xi$ is the
localization length of the QD wavefunctions. In contrast, the interband transition matrix element between the two continua
does not have such exponential suppression. Thus, with the help of the continua, the photon-assisted charge transfer 
between the two QDs layers is significantly increased.

\section{Linear-response theory}
In the linear-response regime, we can characterize thermoelectric transport in the device using a simple 
current model. As shown in Refs.~\onlinecite{nf0,nftpv0}, in near field heat transfer, the temperature gradients
in the receiving material due to inhomogeneous heating is negligible. For simplicity, we assume that electrons
in the two continua give energy to phonons and quickly reach thermal equilibrium with the substrate. Hence the
electronic temperatures in the two continua and in the source and the drain are assumed to be the same, which is denoted as
$T_c$. However, since the thermal radiation is continuously
pumping electrons from the valence band to the conduction band, the ``pseudo" electrochemical
potentials of these two continua are different from the equilibrium chemical potential. We denote the ``pseudo" electrochemical
potential of the valence (conduction) band as $\mu_v$
($\mu_e$). Consider the three processes which are described by three
electrical currents: the electron flow from the source to the valence
band, $I_{S,v}$; the electron transition between the valence band and
the conduction band, 
described as a current via $I_{v,e}=e\Gamma_{ve}$, where $\Gamma_{ve}$
is the flux of photo-carriers and $e<0$ is the charge of an electron; the electron flow
from the conduction band to the drain, $I_{e,D}$.
In linear-response, these currents are calculated as,
\begin{subequations}
\begin{align}
& I_{S,v} = G_{\ell} (\mu_S -\mu_v)/e, \label{i1} \\
& I_{v,e} = G_{ve} (\mu_v - \mu_e) /e + L_{ve} (T_h - T_c)/T, \\
& I_{e,D} = G_{r} (\mu_e - \mu_D) /e , \label{i3}
\end{align}
\label{currentEq}
\end{subequations}
where $G_\ell$ and $G_r$ are the electrical conductances of the left and right QD layers, respectively. Here $G_{ve}$ is the conductance
describing the charge transfer between the two continua, while $L_{ve}$ describes the associated Seebeck effect due to heat absorption. 
$\mu_S$ and $\mu_D$ are the electrochemical potentials of the source and the drain, separately. Because the central region is assumed 
to have the same temperature as the source and the drain, the Seebeck effects in (\ref{i1}) and (\ref{i3}) are neglected. Note that the 
interband transition current, $I_{ve}$, is driven by two ``forces": the electrochemical potential difference of the two continua and the temperature
difference between the thermal bath and the device.

In the next section we will present a microscopic theory which
expresses the above transport coefficients using microscopic quantities. For this moment, we focus on the
phenomenological transport properties of the NFI3T heat engine. The three currents in Eq.~(\ref{currentEq}) must be the same to conserve the total charge:
\be
I_{S,v} = I_{v,e} = I_{e,D} .
\ee
These equations determine both $\mu_v$ and $\mu_e$ once $\mu_S$,
$\mu_d$, $T_c$, and $T_h$ are set. In the following, we denote the equilibrium chemical potential and temperature as $\mu$ and $T$,
respectively. Besides, $\mu_S-\mu=-(\mu_D-\mu)=\Delta\mu/2$ with
$\Delta\mu\equiv \mu_S - \mu_D$, while $T_h = T +
\Delta T/2$ and $T_c= T-\Delta T/2$. The solution
of the above equations yields,
\begin{align}
& \mu_e = \frac{( G_{ve}G_\ell - G_{ve}G_r - G_{l} G_r)\Delta\mu/2 + e
 G_\ell L_{ve}\Delta T/T}{G_{l}G_r + G_{ve} G_\ell + G_{ve} G_r} ,\nn \\
& \mu_v = \frac{G_\ell-G_r}{G_\ell}\frac{\Delta\mu}{2} -
\frac{G_r}{G_\ell}\mu_e .
\end{align}
Inserting these solutions, we obtain the following simple results:
\be
  \left( \begin{array}{c}
      I_e\\ I_{Q} \end{array}\right) =
  \left( \begin{array}{cccc}
      G_{eff} & L_{eff} \\
      L_{eff} & K_{ve}
    \end{array}\right) \left(\begin{array}{c}
      \Delta\mu /e \\  \Delta T /T \end{array}\right) .\label{2t-trans}
\ee
Here $I_e$ is the electrical current between the source and the drain, while $I_Q$ is the heat flux transferred from
the thermal terminal to the device. $G_{eff}$ is the total conductance, and  $L_{eff}$ describes the Seebeck effect due to near field
heat transfer. They are given by
\begin{subequations}
\begin{align}
& G_{eff} = (G_\ell^{-1} + G_r^{-1} + G_{ve}^{-1})^{-1} ,\\
& L_{eff} = \frac{G_{eff}}{G_{ve}} L_{ve} ,
\end{align}
\label{master}
\end{subequations}
and $K_{ve}$ is the heat conductance associated with the heat transfer between the thermal terminal and the device.

The above tells us that the figure of merit of our inelastic thermoelectric
device is
\be
ZT = \frac{L_{eff}^2}{G_{eff}K_{tot} - L_{eff}^2}
\ee
where we have included the undesirable parasitic heat conductance $K_{par}$ due to hot-carrier heating and other dissipation effects via 
$K_{tot}=K_{ve}+ K_{par}$. The parasitic heat conduction will be analyzed in details in Sec.~V.

\section{Microscopic theory}

Here we present a microscopic theory for the thermoelectric transport
in the NFI3T heat engines. The Hamiltonian of the system is
\be
H = H_{SD} + H_{QD} + H_C + H_{tun} + H_{e-ph} .
\ee
Here $H_{SD}$, $H_{QD}$, and $H_C$ are the Hamiltonian of the source
and drain, the QDs, and the two central continua,
separately. $H_{tun}$ describes tunneling through the QDs, whereas
$H_{e-ph}$ describe the optical transition between the two continua.
The Hamiltonian for the source and drain is
\be
H_{SD} = \sum_{{\vec q}} (E_{S,{\vec q}} c^\dagger_{S,{\vec q}}
c_{S,{\vec q}}  + E_{D,{\vec q}} c^\dagger_{D,{\vec q}}
c_{D,{\vec q}} ) ,
\ee
where ${\vec q}$ is the wavevector of electrons.
The Hamiltonian of the QDs is
\be
H_{QD} = \sum_{j=\ell,r} E_{j} d^\dagger_j d_j ,
\ee
where $j=\ell,r$ denotes the left and right dot, respectively. We
first consider the case where only one (two if spin degeneracy is
included) level in each QD is relevant for the transport. The
Hamiltonian for the two central continua is 
\be
H_C = \sum_{{\vec q}} ( E_{v,{\vec q}} c^\dagger_{v,{\vec q}}
c_{v,{\vec q}}  +  E_{e,{\vec q}} c^\dagger_{e,{\vec q}} c_{e,{\vec
  q}} ).
\ee
The tunnel coupling through the QDs is given by
\begin{align}
H_{tun} =& \sum_{{\vec q}} ( J_{S,{\vec q}} c_{S,{\vec q}}^\dagger
d_\ell + J_{D,{\vec q}} c_{D,{\vec q}}^\dagger
d_r \nn \\
& + J_{v,{\vec q}} c_{v,{\vec q}}^\dagger
d_\ell  + J_{e,{\vec q}} c_{e,{\vec q}}^\dagger
d_r) + {\rm H.c.} .
\end{align}
The coupling coefficients $J$'s determine the tunnel rates through the
Fermi golden rule,
\begin{subequations}
\begin{align}
& \Gamma_{S,\ell} = \frac{2\pi}{\hbar} \sum_{{\vec q}} |J_{S,{\vec
    q}}|^2\delta(E_\ell-E_{S,{\vec q}}) ,\\
& \Gamma_{D,r} = \frac{2\pi}{\hbar} \sum_{{\vec q}} |J_{D,{\vec
    q}}|^2\delta(E_r-E_{D,{\vec q}}) ,\\
& \Gamma_{v,\ell} = \frac{2\pi}{\hbar} \sum_{{\vec q}} |J_{v,{\vec
    q}}|^2\delta(E_\ell-E_{v,{\vec q}}) ,\\
& \Gamma_{e,r} = \frac{2\pi}{\hbar} \sum_{{\vec q}} |J_{e,{\vec
    q}}|^2\delta(E_r-E_{e,{\vec q}}) .
\end{align}
\end{subequations}
The electrical currents through the QDs are given by ($e<0$ is the electron charge)
\begin{subequations}
\begin{align}
& I_{S,v} = \frac{2e}{h} \int dE {\cal T}_{\ell} (E) [f_S(E) - f_v(E)] ,\\
& I_{e,D} = \frac{2e}{h} \int dE {\cal T}_{r} (E) [f_e(E) - f_D(E)] .
\end{align}
\end{subequations}
Here the factor of two comes from the spin degeneracy (we consider around room-temperature or above,
where Coulomb blockade is negligible). $e$ is the electronic charge. $f_S$, $f_v$, $f_e$, and
$f_D$ are the electron distribution functions in the source, valence
band, conduction band, and drain, separately. The transmission
functions are given by
\begin{subequations}
\begin{align}
& {\cal T}_{\ell} (E) = \frac{\hbar^2\Gamma_{S,\ell}\Gamma_{v,\ell}}{(E-E_\ell)^2 + \frac{\hbar^2}{4}
  (\Gamma_{S,\ell}+\Gamma_{v,\ell})^2} ,\\
& {\cal T}_{r} (E) = \frac{\hbar^2\Gamma_{D,r}\Gamma_{e,r}}{(E-E_r)^2 + \frac{\hbar^2}{4}
  (\Gamma_{D,r}+\Gamma_{e,r})^2} ,
\end{align}
\end{subequations}
The linear-response conductance is given by
\be
{G}_j  =  \frac{2e^2}{hk_BT} \int dE {\cal T}_j (E) f(E)
[1-f(E)], \ \  j=\ell, r . \quad
\ee

{We now consider the photon-assisted transitions in the center. The Hamiltonian governing such transitions is given by
\be
H_{e-ph} = \sum_{{\vec q},{\vec k},\tau} \frac{g_{{\vec k},\tau}}{\sqrt{V}} c^\dagger_{e,{\vec q}+{\vec k}} c_{v,{\vec q}} a_{{\vec k},\tau} +
{\rm H.c.},
\ee
where $g_{\vec k}$ is the electron-photon interaction strength, the operator $a_{{\vec k},\tau}$ 
($\tau=s,p$ denotes the $s$ and $p$ polarized light) annihilates
an infrared photon with polarization $\tau$. $V$ is the volume of the photonic system. The electron-photon interaction strength is 
determined by the  inter-band dipole matrix elements and the photon frequency $\ome_{\vec k}$. For instance, the electron-photon interaction
in a direct-gap semiconductor is 
\be
g_{{\vec k},\tau} = i \sqrt{\frac{\hbar\ome_{\vec k}}{2\vep_0\vep_r }} d_{cv} ,
\ee
where $\vep_0$ and $\vep_r$ are the vacuum and relative permittivity, respectively, and $d_{cv}$ is the inter-band dipole matrix element.
The Fermi golden rule determines the electrical current generated by inter-band transitions,
\begin{align}
I_{ve} =& \frac{2\pi e}{\hbar} \sum_{{\vec k}, {\vec q},\tau} \frac{|g_{{\vec k},\tau}|^2}{V}
\delta(E_{e,{\vec k}+{\vec q}}-E_{v,{\vec q}}-\hbar\ome_{{\vec k}}) \nn\\
&\times \{f_v(E_{v,{\vec q}})[1-f_e(E_{e,{\vec k}+{\vec q}})]\tilde{N}_{{\vec k},\tau} \nn \\
&\ \quad - f_e(E_{e,{\vec k}+{\vec q}})[1-f_v(E_{v,{\vec q}})](\tilde{N}_{{\vec k},\tau}+1)\} ,
\end{align}
where $f_v$ and $f_e$ are the electronic distribution functions for the lower
and upper continua, respectively. The nonequilibrium photon distribution
\begin{align}
& \tilde{N}_{{\vec k},\tau} = N^0(\ome_{\vec k}, T_c) + \delta {N}_{{\vec k},\tau},\nn\\
&  \delta {N}_{{\vec k},\tau}= [N^0(\ome_{\vec k}, T_h) -N^0(\ome_{\vec k}, T_c)]{\cal T}_{\tau}(\ome_{\vec k}, k_\parallel, d) 
\end{align}
consists of the equilibrium photons in the absorption material $N^0(\ome_{\vec k}, T_c)=1/[\exp(\frac{\hbar\ome_{\vec k}}{k_BT_c})-1]$ 
and the hot photons tunneled from the thermal emitter. The probability of encountering such hot photons is determined by the product of 
their distribution and the tunneling probability. The photon tunneling probability between the emitter and the absorber across a planar 
vacuum gap is a function of photon frequency $\ome_{\vec k}$, the amplitude of the wavevector parallel to the planar interface 
$k_\parallel=|{\vec k}_\parallel|$, and the thickness of the gap $d$ \cite{nf-1,nftpv3},
\begin{align}
{\cal T}_{\tau}(\ome_{\vec k}, k_\parallel, d) = \left\{ \begin{array}{cccc} \frac{(1-|r^\tau_{01}|^2)(1-|r^\tau_{02}|^2)}{|1-r^\tau_{01}r^\tau_{02}e^{i2k^0_{z}d} |^2}, \quad {\rm if} \quad k_\parallel \le \ome/c \\
\frac{4\Im(r^\tau_{01})\Im(r^\tau_{02})e^{-2\beta^0_z d} }{|1- r^\tau_{01}r^\tau_{02}e^{-2\beta^0_z d} |^2} , \quad {\rm otherwise}
\end{array} \right.
\label{pT}
\end{align}
Here $r^\tau_{01}$ ($r^\tau_{02}$) is the Fresnel reflection coefficient for the interface 
between the vacuum (denoted as ``0") and the emitter (absorber) [denoted as ``1" (``2")]. $k^0_z=\sqrt{(\ome/c)^2-k_\parallel^2}$ is 
the wavevector perpendicular to the planar interfaces in the vacuum. For $k_\parallel>\ome/c$, the perpendicular wavevector
in the vacuum is imaginary $i\beta^0_z=i\sqrt{k_\parallel^2-(\ome/c)^2}$, where photon tunneling is dominated by evanescent waves.
For isotropic electromagnetic media, the Fresnel coefficients are given by \cite{nftpv3}
\begin{subequations}
\begin{align}
& r^s_{0j} = \frac{k_z^0-k_z^j}{k_z^0+k_z^j}, \\
& r^p_{0j} = \frac{\vep_jk_z^0 - k_z^j}{\vep_jk_z^0 + k_z^j} , \quad j=1,2 ,
\end{align}
\end{subequations}
where $k_z^j=\sqrt{\vep_j(\ome/c)^2-k_\parallel^2}$ and $\vep_j$ ($j=0,1,2$) are the (complex) wavevector along the $z$ direction
and the relative permittivity in the vacuum, emitter, and the absorber, separately.}

From the above equations we obtain the linear thermoelectric transport
coefficients,
\begin{subequations}
\begin{align}
& G_{ve} = \frac{e^2}{k_BT}\int d\ome \Gamma_0(\ome) , \\
& L_{ve} = \frac{e}{k_BT}\int d\ome \Gamma_0(\ome) \hbar \ome ,\\
& K_{ve} = \frac{1}{k_BT}\int d\ome \Gamma_0(\ome) \hbar^2 \ome^2 . 
\end{align}
\label{ave-E} 
\end{subequations}
where 
\begin{align}
 \Gamma_0(\ome) =& 2\pi \nu_{ph} {\cal F}_{nf} (\ome) \sum_{{\vec q}} |g(\ome)|^2
\delta(E_{e,{\vec q}}-E_{v,{\vec q}}-\hbar\ome) \nn\\
&\times f^0(E_{v,{\vec q}}, T)[1-f^0(E_{e,{\vec q}}, T)]N^0(\ome, T) .  \label{gam0}
\end{align}
{The superscript 0 in the above stands for the equilibrium distribution, and $f^0(E_{v,{\vec q}}, T)=1/[\exp(\frac{E_{v,{\vec q}}-\mu}{k_BT})+1]$. $|g(\ome)|^2 = \frac{\hbar\ome d_{cv}^2}{2\vep_0\vep_r}$, 
$\nu_{ph} = \frac{\ome^2\vep_r^{3/2}}{\hbar \pi^2 c^3}$ is the photon density of states, and the factor 
\begin{align}
{\cal F}_{nf}(\ome) &= \frac{1}{\nu_{ph}} \frac{1}{V} \sum_{{\vec k},\tau} \delta (\hbar\ome - \hbar\ome_{\vec k}) {\cal T}_\tau(\ome, k_\parallel, d),\nn \\
&= \frac{1}{4} \int_0^{1} \frac{x_kdx_k}{\sqrt{1-x_k^2}} \sum_\tau {\cal T}_{\tau} (\ome, x_k n \ome/c, d) .
\end{align}
where $x_k=k_\parallel/(n\ome/c)$. In the above, we have used the fact that the wavevector of photons are much smaller than that of electrons,
thus ${\vec q}+{\vec k}\simeq {\vec q}$. In generic situations, the above transition rate also depends on 
the position (e.g., if the local density of states of photon is not uniform due to, say, surface plasmon polaritons), and
an integration over the whole central region is needed. For simplicity, we consider for now only propagating photons in the absorption material}. 
The Seebeck coefficient of the NFI3T heat engine is
\be
S = \frac{L_{eff}}{T G_{eff}} = \frac{\ave{\hbar\ome}}{eT} ,
\ee 
where the average is defined as 
\be
\ave{...} = \frac{\int d\ome \Gamma_0(\ome) ... }{\int d\ome
  \Gamma_0(\ome)} .
\ee
We can then express the figure of merit in microscopic
quantities: 
\begin{align}
ZT = \frac{\ave{\hbar\ome}^2}{\alpha\ave{\hbar^2\ome^2}-\ave{\hbar\ome}^2+\Lambda_{nf}} , \label{ztnf}
\end{align}
where $\alpha = \frac{G_{ve}}{G_{eff}}$, and $\Lambda_{nf}=e^2K_{para}/G_{ve}$ characterizes the parasitic heat conductance
$K_{para}$ that does not contribute 
to thermoelectric energy conversion. {Although $\alpha$ is always greater than unity, in this work we shall focus on
the regime with $\alpha \simeq 1$, which can be realized by using highly conducting layers on the left and right sides of
the absorption material. Beside high-density quantum-dots layers,  quantum wells or doped semiconductors
can also be used to attain high conductivity between the source (drain) and the absorption material. In addition, at this stage
we shall not quantify the parasitic heat conduction that does not contribute to thermoelectric energy conversion, i.e., we 
set $\Lambda_{nf}=0$ and hence the heat current $I_Q$ counts only absorbed infrared photons. We will discuss 
briefly possible parasitic heat conduction mechanisms in Sec.~VC and the effects of parasitic heat conduction on thermoelectric 
performance. We remark that by setting $\Lambda_{nf}=0$ and considering only photons with $\hbar\ome>E_g$, 
we assume 100\% recycling of low-frequency $\hbar\ome<E_g$ and unabsorbed photons
(e.g., they can be reflected back to the emitter by a ``back-side reflector" and be thermalized with other quasi-particles in the emitter again).
Therefore, the only way that heat leaves the hot bath is by exciting electrons from the lower to the upper band via light-matter
interaction.}

We now estimate the conductances $G_\ell$, $G_r$, and $G_{ve}$. The tunneling conductances are estimated as
\begin{align}
& G_\ell \simeq \frac{2e^2}{h} n_{QD} f_\ell^0 (1-f_\ell^0) \frac{\Gamma_\ell}{k_BT} , \\
& G_r \simeq \frac{2e^2}{h} n_{QD} f_r^0 (1-f_r^0) \frac{\Gamma_\ell}{k_BT} , \label{estimation}
\end{align}
when $\Gamma$ is smaller than $k_BT$. Here $f_\ell^0=n_F(E_\ell,T)$ and $f_r^0=n_F(E_r,T)$ where $n_F$ is the equilibrium Fermi
distribution, and $n_{QD}$ is the density of QDs which can be as large as $10^{15}$~m$^{-2}$ (see, e.g., Ref.~\onlinecite{qd}). 
These conductances can reach about 100~S/m. If the thickness of the QD layer is 10~nm, the conductance is about $10^{10}$~S.

{The conductance associated with the inter-band transition can be roughly estimated as}, 
\be
G_{ve}\sim \frac{e^2\delta^\prime}{k_BT}  \rho_{ve} |g|^2 {\cal F}_{nf} \nu_{ph}(\ave{\hbar\ome}) N^0(\ave{\hbar\ome}, T) .
\ee
Here $\delta^\prime={\rm min}(k_BT,\delta_{ve},\delta_{ph})$ where $\delta_{ve}$ is the band width of the continua, $\delta_{ph}$ is the 
bandwidth of the near-field photons, $\rho_{vc}$ is the joint electronic 
density of states for the lower and upper continua, and $\ave{\hbar\ome}$ is the average photon energy for photon-assisted interband transitions.
We emphasize that, {in comparison with photon-assisted hopping between QDs considered before \cite{3tp1,3tp2}, here the conductance
is much enhanced} because of the following reasons: (1) As stated before, the electron-photon interaction is not
reduced by the ``form factor" of the QDs as \cite{3t11} $\sim \exp(-2d/\xi)$; (2) The joint-density-of-states of the two continua can be  
larger than that of QD-ensembles.

\begin{figure}[htb]
  \centering \includegraphics[width=4.cm]{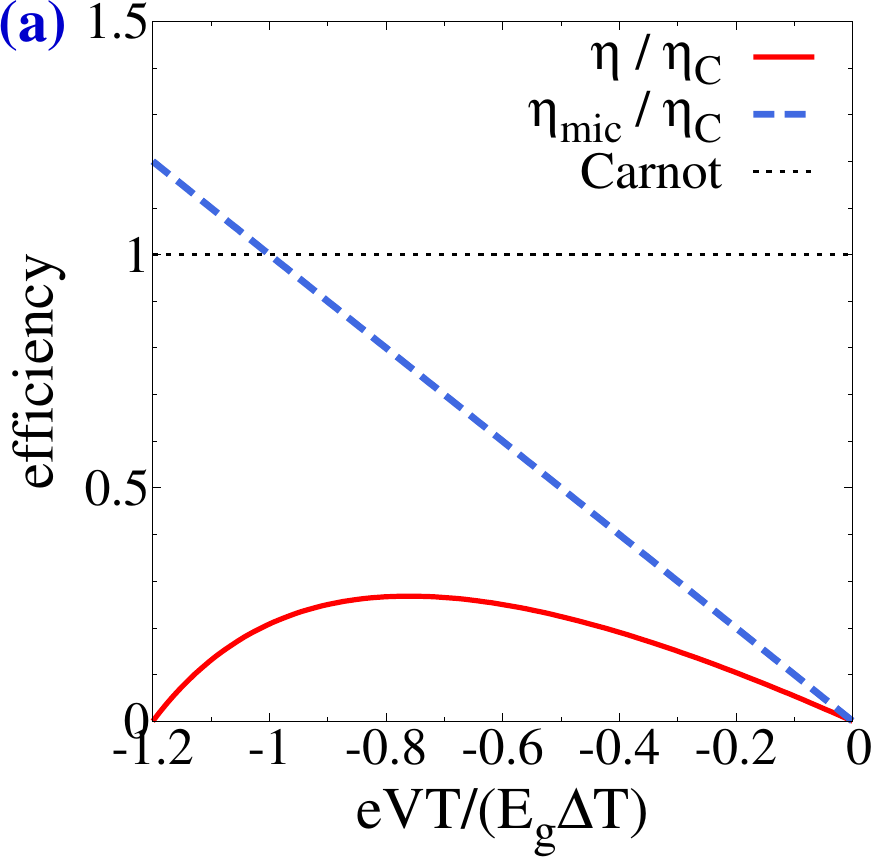}\hspace{0.2cm}\includegraphics[width=4.cm]{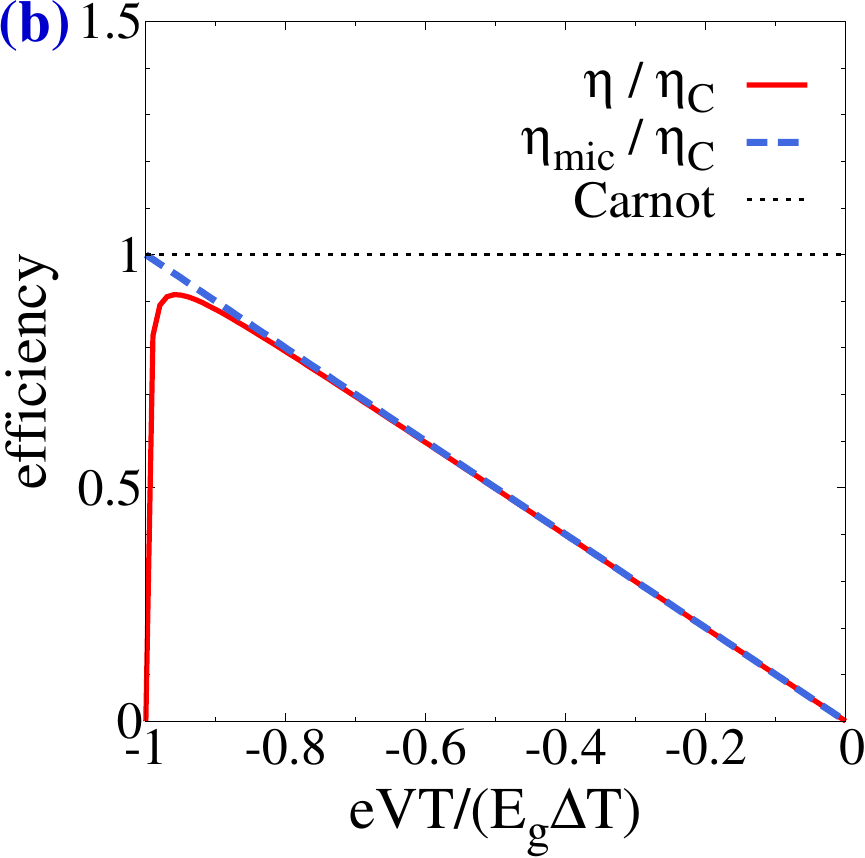}
  \caption{(Color online) Energy efficiency of the NFI3T heat engine vs. the voltage bias $V$ for (a) $\alpha=1$, $\beta=0.5$, and
  $\ave{x_\ome}=1.2$, and (b) $\alpha=1$, $\beta=0.002$, and $\ave{x_\ome}=1$. The calculation of the efficiencies are according to 
  Eqs.~(\ref{effa}) and (\ref{effb}). The energy efficiencies are scaled by the Carnot efficiency.  }
\label{fig2}
\end{figure}

\section{Thermoelectric Performance}

\subsection{General analysis}
In a simple picture, each infrared photon (say, with energy $\hbar\ome=E_g$) absorbed by the NFI3T device is used to move an 
electron from the source to the drain. {The energy quantum for the absorbed photon is $\hbar\ome\gtrsim E_g$, while the energy 
increment of the electron is $|eV|$, if there is a voltage bias $V=(\mu_S-\mu_D)/e$ ($\mu_D>\mu_S$) between the source and 
the drain. We then define a ``microscopic" energy efficiency,}
\be
\eta_{mic} = \frac{|eV|}{E_g}  ,\label{etamic}
\ee
{which characterizes the ratio of the output energy over the input energy for each photon absorption processes.}
At first sight, it seems that the efficiency can be improved by applying a larger voltage bias. However, when detailed-balance effects 
are taken into account, the efficiency is reduced. First, the back flow of the electron
and the emission of photon by recombination of carriers are also possible. The net current from source to drain is determined by 
thermodynamic balance of the forward (electricity generation) and backward (carrier recombination and photon generation) processes. Besides, 
the photon source is not monochromatic, but follows the Bose-Einstein distribution. Since the increase of photon energy reduces
the efficiency, the broad energy distribution of photons  also reduces the efficiency. Also, there are photons that are not absorbed (reflected or transmitted).
The first and the third mechanisms are the main factors that reduce the efficiency.

To give a simple quantitative picture, we show how the efficiency varies with the voltage bias. To characterize the thermoelectric transport coefficients, 
we introduce the dimensionless parameters, $u\equiv eVT/E_g\Delta T$, $\ave{\hbar^2\ome^2}\equiv (1+\beta)\ave{\hbar\ome}^2$, $x_\ome\equiv \hbar\ome/E_g$. Using 
Eq.~(\ref{ave-E}), the thermoelectric transport can be written as
\be
  \left( \begin{array}{c}
      I_e\\ I_{Q} \end{array}\right) =
  G_{eff} \left( \begin{array}{cccc}
     1 & \ave{\hbar\ome}/e \\
     \ave{\hbar\ome}/e & \alpha \ave{\hbar^2\ome^2}/e^2
    \end{array}\right) \left(\begin{array}{c}
     V \\  \Delta T /T \end{array}\right) .\label{2t-trans}
\ee
where $G_{eff}$ is the total conductance of the device. The energy efficiency can then be written as 
\begin{align}
\frac{\eta}{\eta_C} = \frac{-u(u+\ave{x_\ome})}{u\ave{x_\ome}+\alpha(1+\beta)\ave{x_\ome}^2} .\label{effa}
\end{align}
The energy efficiency of the heat engine is always smaller than the Carnot efficiency $\eta_C\equiv \Delta T/T$, 
since $\alpha> 1$ and $\beta>0$. The working region of the heat engine is $-\ave{x_\ome}<u<0$. On the other hand, we have
\begin{subequations}
\begin{align}
& \frac{\eta_{mic}}{\eta_C} = - u, \\
& \frac{\eta}{\eta_{mic}} = \frac{u+\ave{x_\ome}}{u\ave{x_\ome}+\alpha(1+\beta)\ave{x_\ome}^2} . 
\end{align}
\label{effb}
\end{subequations}
In Fig. 2 we study two examples: (i) $\alpha=1$, $\beta=0.5$, $\ave{x_\ome}=1.2$ [Fig.~2(a)], and 
(ii) $\alpha=1$, $\beta=0.002$, $\ave{x_\ome}=1$ [Fig.~2(b)]. The first example is closer to realistic 
devices where the optimal efficiency is considerably smaller than the Carnot efficiency, due to dissipations. From Fig. 2a one can see 
that the efficiency initially increase with the voltage. However, the back flow of 
electrons becomes more and more important as the voltage increases. After an optimal value, the efficiency then decreases with the voltage. 
Consistent with the second law of thermodynamics \cite{3tpre}, the optimal efficiency is always smaller than the Carnot efficiency. The second 
example  has a much smaller dissipation (in fact, it is very close to the reversible limit). In a quite large range, the efficiency indeed increases 
with the voltage and is almost close to the microscopic efficiency $\eta_{mic}$ [see Fig.~2(b)]. The ratio 
$\frac{\eta}{\eta_{mic}}$ is always a decreasing function of $|u|=|eVT/E_g\Delta T|$. {However, as $u$ reaches close to 1, the energy efficiency 
rapidly reduces from the microscopic efficiency $\eta_{mic}$, staying smaller than the Carnot efficiency. We remark that with perfect energy filters,
a heat engine that reaches the Carnot efficiency can be obtained \cite{kedem,3tp1,prb1,low-diss,3t7,3tp3}.}
 
\begin{figure}[htb]
  \centering \includegraphics[width=4.2cm]{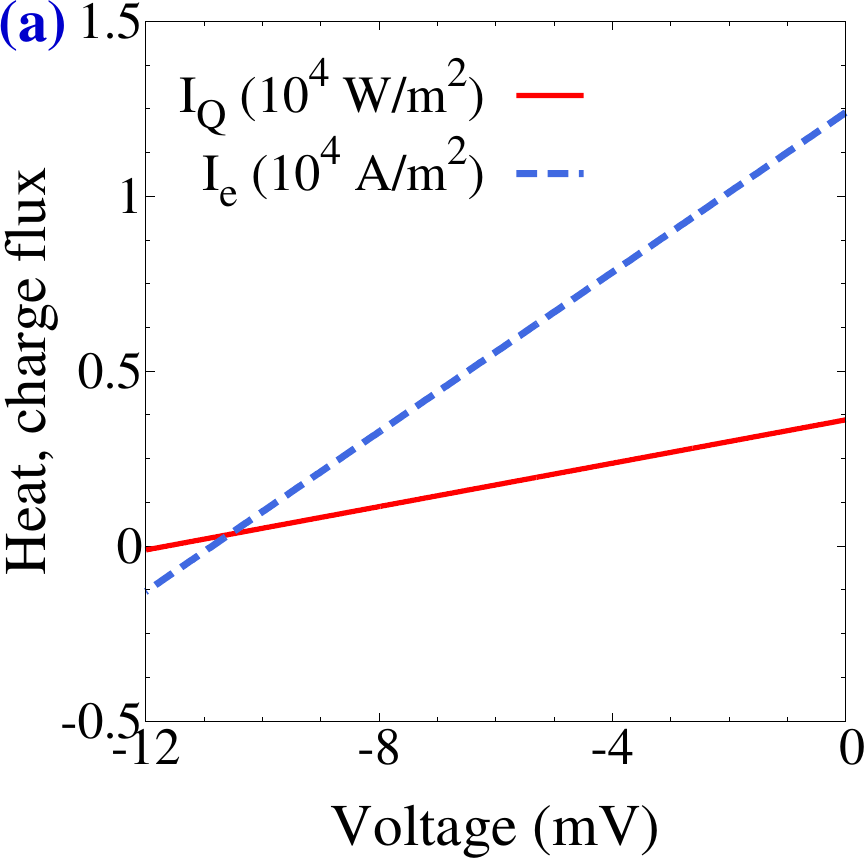}\hspace{0.2cm}\includegraphics[width=4.2cm]{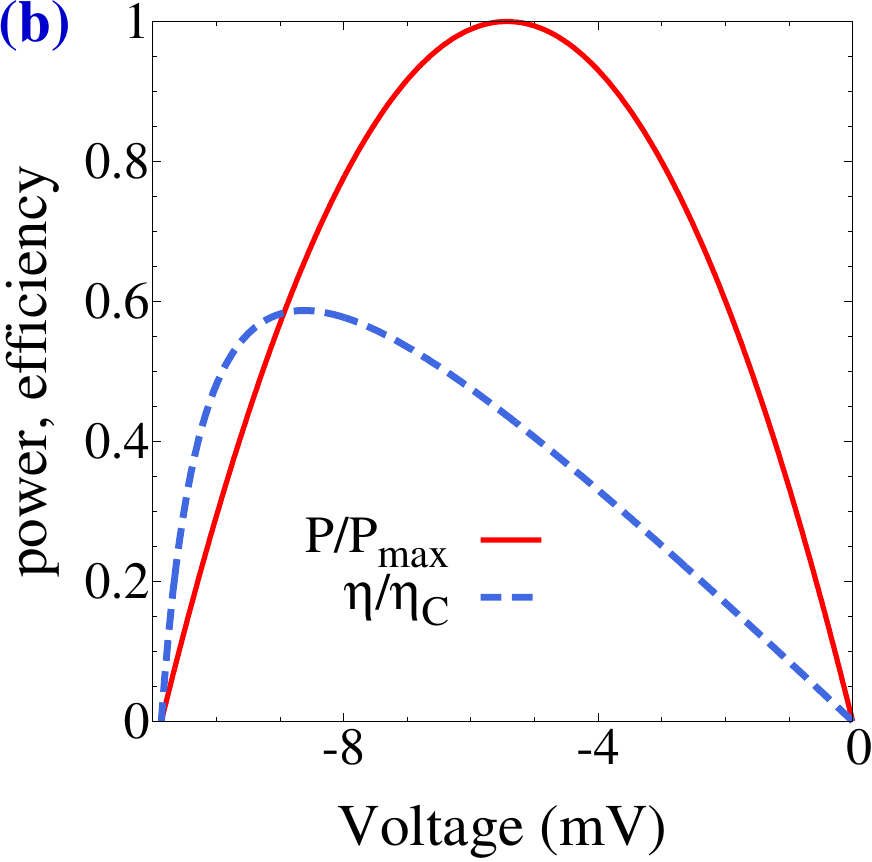}
  \caption{(Color online)  (a) Heat and charge fluxes in the NFI3T heat engine as functions of the voltage $(\mu_S-\mu_D)/e$ for 
  $T_c=500$~K, $\mu=0.1$~eV, and $\Delta T=20$~K. $I_{Q}$ is the absorbed heat flux, while $I_{e}$ is the induced charge current.
  (b) The output power $P=-I_eV$ and the energy efficiency $\eta$ of the heat engine for the same conditions. Here the maximum 
  output power for this condition is $P_{max}=33.7$~W/m$^2$. The two electronic continua in the center are the valence and conduction 
  bands of InSb, which has a direct band gap of $E_g=0.17$~eV. {The thickness of the vacuum gap is $d=100$~nm.}}
  \label{fig3}
\end{figure}

\subsection{Calculation of thermoelectric transport and performance using InSb as absorption material}

We now present a more concrete calculation of the electrical conductivity, thermopower, the figure of merit, and the power factor
for a particular NFI3T heat engine where the two continua in the center are the valence and conduction bands of a narrow band 
gap semiconductor, InSb. The band gap of InSb is 0.17 eV. To avoid further complication, we do not consider the surface electromagnetic waves and
temperature-dependence of the band gap. In this situation, the electron-photon interaction coefficient is given by
\be
|g(\ome)|^2 = \frac{\hbar\ome d_{cv}^2}{2\vep_0\vep_r}, \label{insb1}
\ee
where $d_{cv}=8.8\times 10^{-28}$~C.m is the interband dipole matrix element in InSb \cite{kp}, $\vep_0$ is the vacuum permittivity, 
$\vep_r=15.7$ is the (high-frequency) dielectric constant, and $c$ is the 
speed of light in vacuum. The dispersion of the conduction and valence band are given by 
\be
E_{v,{\vec k}} = -\frac{\hbar^2k^2}{2m_v}, \quad E_{e,{\vec k}} = E_g + \frac{\hbar^2 k^2}{2m_e} , \label{insb2}
\ee
where the effective masses are $m_e=0.0135m_0$ and $m_v=0.43m_0$ with $m_0$ being the mass of free electron.
We study a thin film of $l_{ab}=10^{-6}$~m thickness sandwiched by the front (source) and back (drain) gates. The thermoelectric efficiency and 
power are calculated using the above equations (for details, see Appendix A). The thermoelectric transport coefficients are calculated using 
Eqs.~(\ref{master}) and (\ref{ave-E}).
We also take into account the conductance of the conduction and valence electrons using the formulae for conductivity, $\sigma_e=n|e|\mu_e$
and $\sigma_v=p|e|\mu_p$. Here $n$ and $p$ are the electron and hole densities, respectively, while $\mu_e$ and $\mu_p$ are the electron
and hole mobilities, respectively. The electron and hole densities are calculated using the Fermi distribution of electrons once the steady-state
chemical potentials and temperatures are given. We use the mobilities of electrons and holes from empirical temperature-dependences obtained from 
experimental data on intrinsic InSb \cite{n,p}: $\mu_p=5.4\times 10^{2}/T^{1.45}$~m$^2$/Vs and $\mu_e=3.5\times 10^{4}/T^{1.5}$~m$^2$/Vs 
where the temperature $T$ is in units of Kelvin. The total conductance is given by $G_{eff}^{-1}=G_\ell^{-1}+G_v^{-1} + G_{ve}^{-1} + G_{e}^{-1} 
+ G_r^{-1}$, where $G_e=\sigma_e/l_{ab}$ and $G_{v}=\sigma_v/l_{ab}$ are the conductances for the upper and lower continua, respectively, 
and the conductances of the left and right QDs layers are taken as $G_\ell=G_r=10^{10}$~S for 10~nm layers [as from the estimation after Eq.~(\ref{estimation})], respectively. We find that
the total conductance is in fact mainly limited by the optical absorption $G_{ve}$. Thus, the conductances of the other parts are not important 
and $\alpha$ is close to unity. For simplicity, we assume that the emitter has the same permittivity of $\vep_r=15.7$ as the absorber.
The temperature range considered here is between 100~K to 600~K, while the melting temperature of InSb is about 800~K. 
{We focus on small voltage and temperature biases, while linear-response theory is applicable.}

\begin{figure}[htb]
  \centering \includegraphics[width=4.2cm]{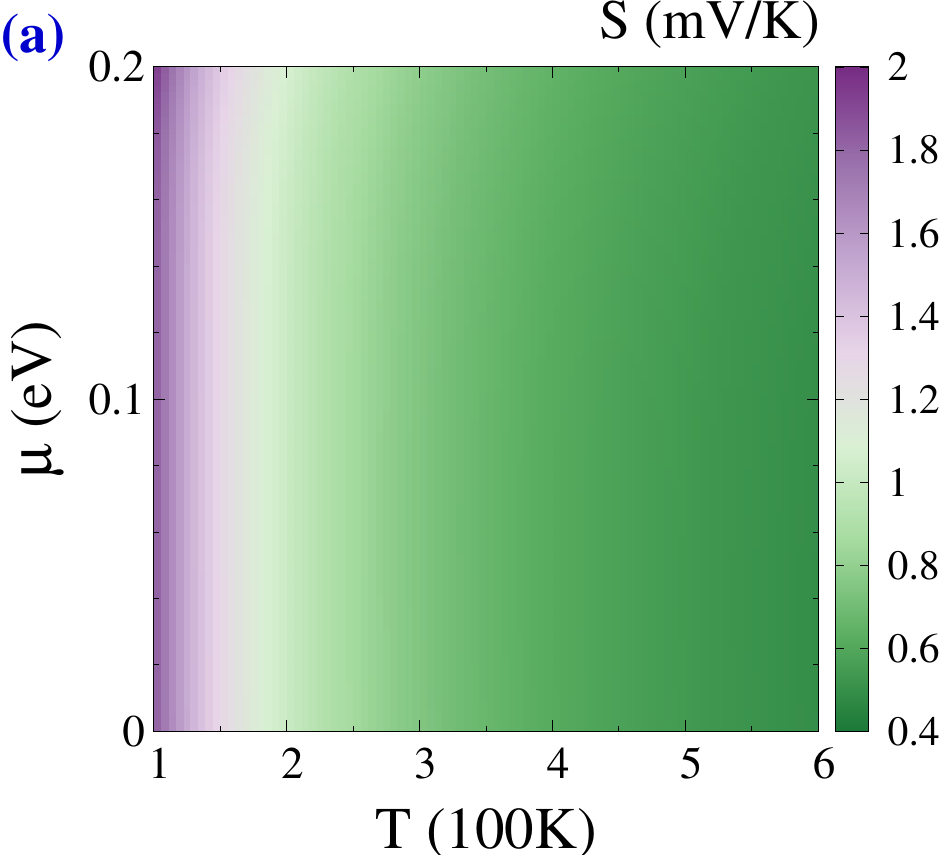}\hspace{0.2cm}\includegraphics[width=4.2cm]{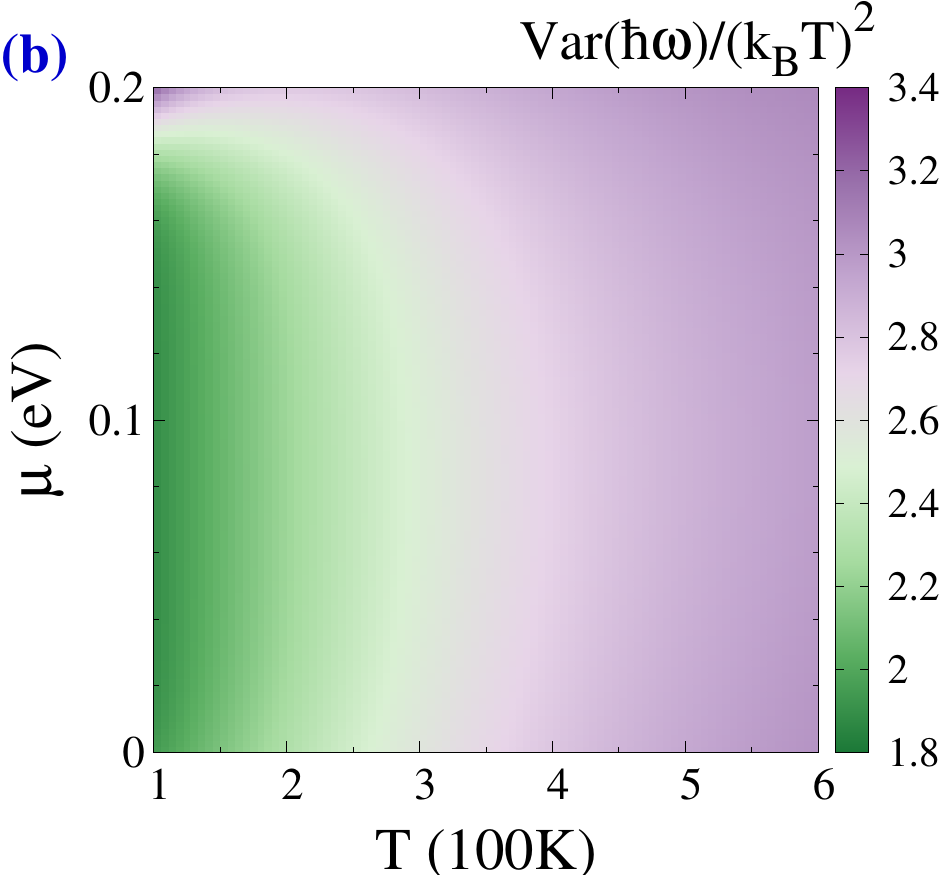}
  \centering \includegraphics[height=4.cm]{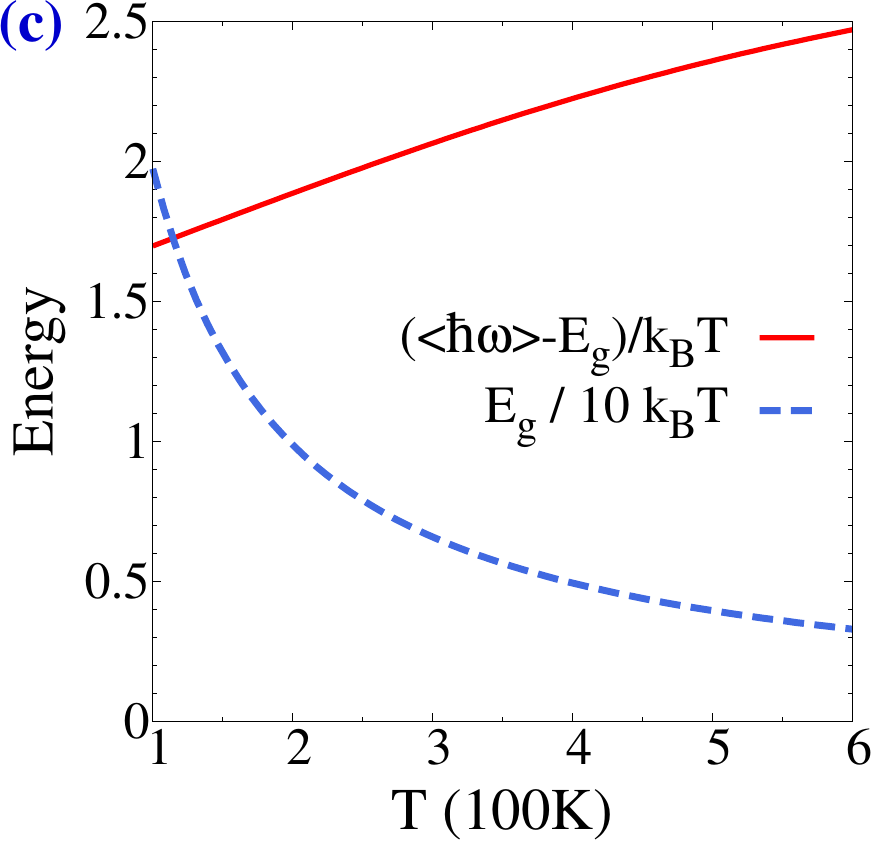}\hspace{0.2cm}\includegraphics[height=4.cm]{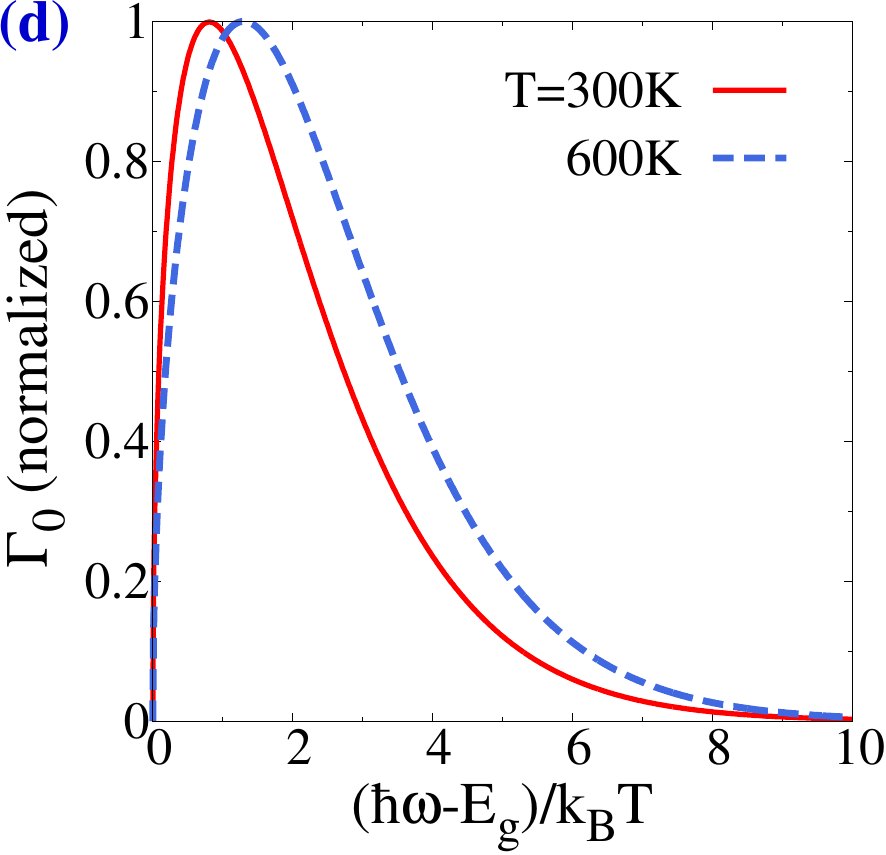}
  \caption{(Color online) (a) Seebeck coefficient $S$ (in unit of mV/K) and (b) the variance of the photon energy ${\rm Var}(\hbar\ome)/k_B^2T^2$
  for the inelastic thermoelectric transport as functions of the chemical potential $\mu$ and the temperature $T$. The average is weighted 
  by the conductance for each channel as in Eq.~(\ref{ave-E}).  The interband transition rate is calculated using Eqs.~(\ref{gam0}), 
  (\ref{insb1}) and (\ref{insb2}). The energy zero is set to be the band edge of the valence band of InSb. (c) The typical energy scales in the
  inelastic thermoelectric transport versus the temperature $T$: the average excess kinetic energy $\ave{\hbar\ome}-E_g$ measured in $k_BT$ 
  and the ratio $E_g/k_BT$. Note that to scale the two together, the latter is divided by a factor of 10 (i.e., the band gap is much larger than $k_BT$). 
  (d) The normalized function $\Gamma_0(\ome)$ as a function of the excess kinetic energy $\hbar\ome-E_g$ (``normalized" in the sense that the 
  maximal values in the figure is 1) for two different temperatures, {$T=300$ and 600~K, respectively. The equilibrium chemical potential 
  for (c) and (d) is $\mu=0.1$~eV. The thickness of the vacuum gap is $d=100$~nm.}}
\label{fig4}
\end{figure}

{In Fig.~3 we plot the heat and charge fluxes, the output electrical power, and the energy efficiency as functions of the applied voltage for 
a NFI3T heat engine with equilibrium chemical potential $\mu=0.1$~eV. The temperatures are $T_h=520$~K and $T_c=500$~K. Using InSb as the
light absorption material, we find that the optimal efficiency can reach to 60\% of the Carnot efficiency. The maximum output power is as large as $34$~W/m$^2$. The absorbed heat flux reaches $0.38\times 10^4$~W/m$^2$.}

{The Seebeck coefficient $S=\ave{\hbar\ome}/eT$ as a function of the chemical potential and temperature is plotted in Fig.~4(a). It is seen that 
the Seebeck coefficient  does not vary considerably with the chemical potential, which is a characteristic of the inelastic thermoelectric
effect, since the average energy $\ave{\hbar\ome}$ is mainly limited by the band gap $E_g$ and the temperature. The variance of the
energy ${\rm Var}(\hbar\ome)\equiv \ave{\hbar^2\ome^2}-\ave{\hbar\ome}^2$ is plotted in Fig.~4(b). From the figure one can see that the variance
of the energy is close to $(k_BT)^2$, which is consistent with the physical intuition. We remark that for the whole temperature and chemical potential
ranges, the variance ${\rm Var}(\hbar\ome)\lesssim 4(k_BT)^2$ is much smaller than the square of the average energy $\ave{\hbar\ome}^2\sim 
100(k_BT)^2$. Particularly for low temperatures, the ratio $\ave{\hbar\ome}^2/{\rm Var}(\hbar\ome)$ can be as large as $\sim 200$. From this
observation, we predict that the figure-of-merit of the NFI3T heat engine can be very large.}

{Fig.~4(c) shows that for various temperatures the average excess kinetic energy of photo-carriers is around $2k_BT$. The average of 
energy of the absorbed photons is mainly determined by the band gap $E_g$. Since $E_g$ is much larger than the thermal energy $k_BT$
[also shown in Fig.~4(c)], the Seebeck coefficient is very large. To give a straightforward understanding of the
energy-dependence of the photon absorption rate, we plot $\Gamma_0(\hbar\ome)$ for two temperatures, $T=300$~K and 600~K, respectively,  
with $\mu=0.1$~eV, in Fig.~4(d). The sharp peaks in the figure manifest efficient energy filtering in the inelastic transport processes.}

\begin{figure}[htb]
  \centerline{ \includegraphics[width=4.2cm]{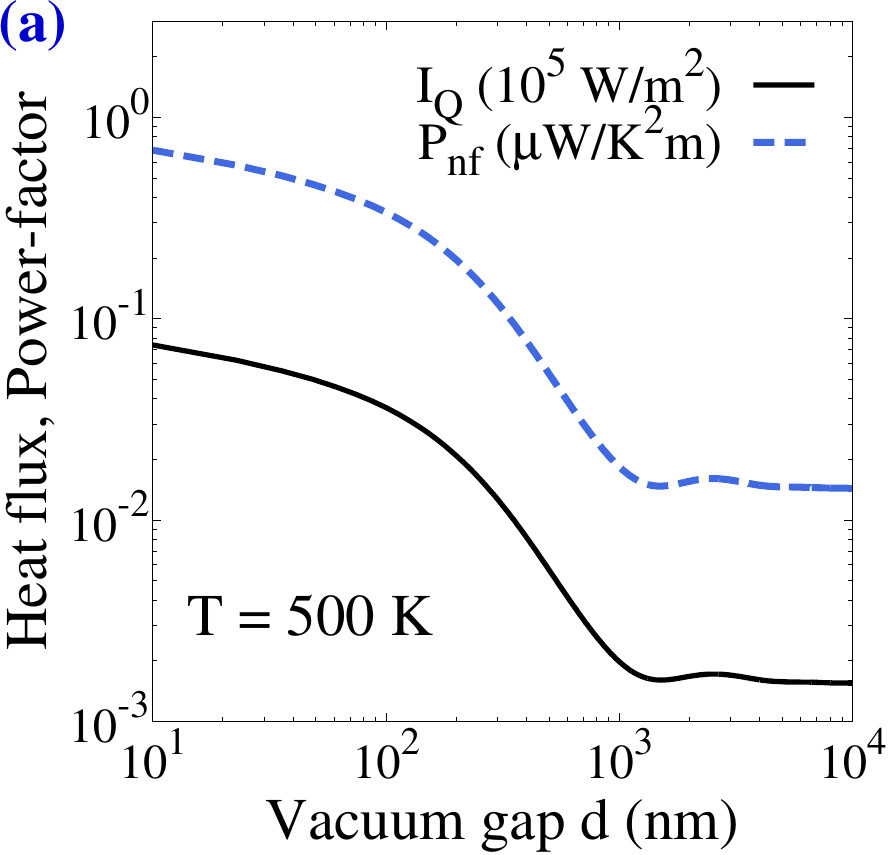}\hspace{0.2cm}\includegraphics[width=4.2cm]{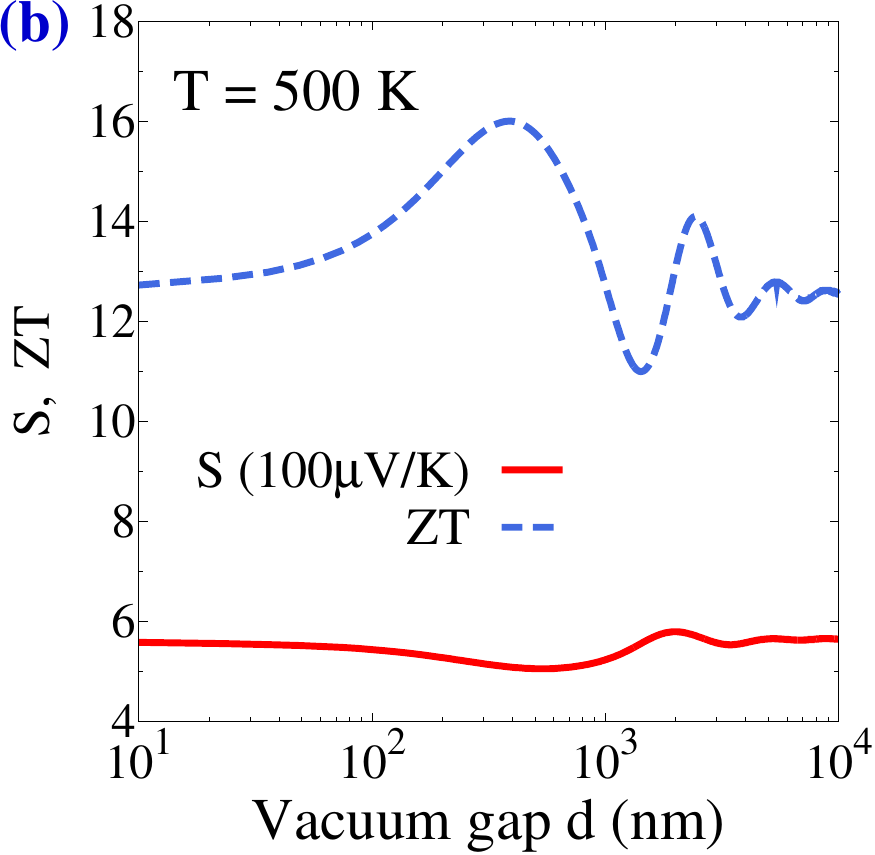}}
   \caption{(Color online) (a) Absorbed heat flux $I_Q$ and power factor $P_{nf}=\sigma_{eff}S^2$ as functions of the vacuum gap $d$. (b) Seebeck 
   coefficient $S$ and thermoelectric figure of merit $ZT$ as functions of the vacuum gap $d$. Equilibrium temperature $T=500$~K and chemical 
   potential $\mu=0.1$~eV. The heat flux is calculated at zero voltage bias and $\Delta T$=20~K.}
\label{fig5}
\end{figure}

{To demonstrate the near-field effect, we calculate the heat flux, the power factor, Seebeck coefficient, and thermoelectric 
figure-of-merit as functions
of the vacuum gap $d$. The results are presented in Fig.~5 where the temperature and chemical potential are set as $T=500$~K and $\mu=0.1$~eV, respectively. We find that the heat-flux and the power factor are significantly enhanced by the near-field effect when the vacuum gap $d$ is decreased
from 10~$\mu$m to 10~nm. This result also indicates that the heat and charge conductivity are significantly improved by the near-field effect.
The Seebeck coefficient and the figure-of-merit, however, exhibit oscillatory dependences on the vacuum gap $d$. The oscillation of thermoelectric
figure-of-merit originates from the oscillation of the Seebeck coefficient $\ave{\hbar\ome}/eT$ and the variance of photon energy ${\rm Var}(\hbar\ome)$.
These oscillations are commonly found in near-field photon transmission, since the interference effects that strongly modulate the photon transmission
have periodic dependence on the vacuum gap $d$, as shown in Eq.~(\ref{pT}).}

\subsection{Parasitic heat conduction and optimization of thermoelectric performance}
{Here, parasitic heat conduction represents the part of heat radiation 
absorbed by the device but not contributed to thermoelectric energy conversion.
Besides the unabsorbed photons from the thermal terminal (which can be reflected back by placing a mirror on the back-side of the 
device), there are other mechanisms for parasitic heat conduction. First, via photon---optical-phonon interactions, photons can directly 
give energy to phonons. Such photon-phonon conversion leads to unwanted heating of the device. Second, photo-carriers can give energy to 
phonons by non-radiative recombinations (e.g., via Shockley-Hall-Read mechanism). Third, hot-carriers can emit phonons and give energy to 
the lattice. We suspect that the first mechanism is rather weak, while the second mechanism depends on the density of impurities and disorder 
in the absorption material. The third mechanism is usually considered as one of the main mechanism for parasitic
heat conduction and reduction of efficiency. We remark that the third mechanism has already been taken into account in our calculation.
According to Fig.~4(b), such parasitic heat conduction leads only to small parasitic heat $\lesssim 2k_BT$ per photo-carrier, which is the
main reason for the high figure-of-merit of the NFI3T heat engine.}

\begin{figure}[htb]
  \centerline{ \includegraphics[height=4.1cm]{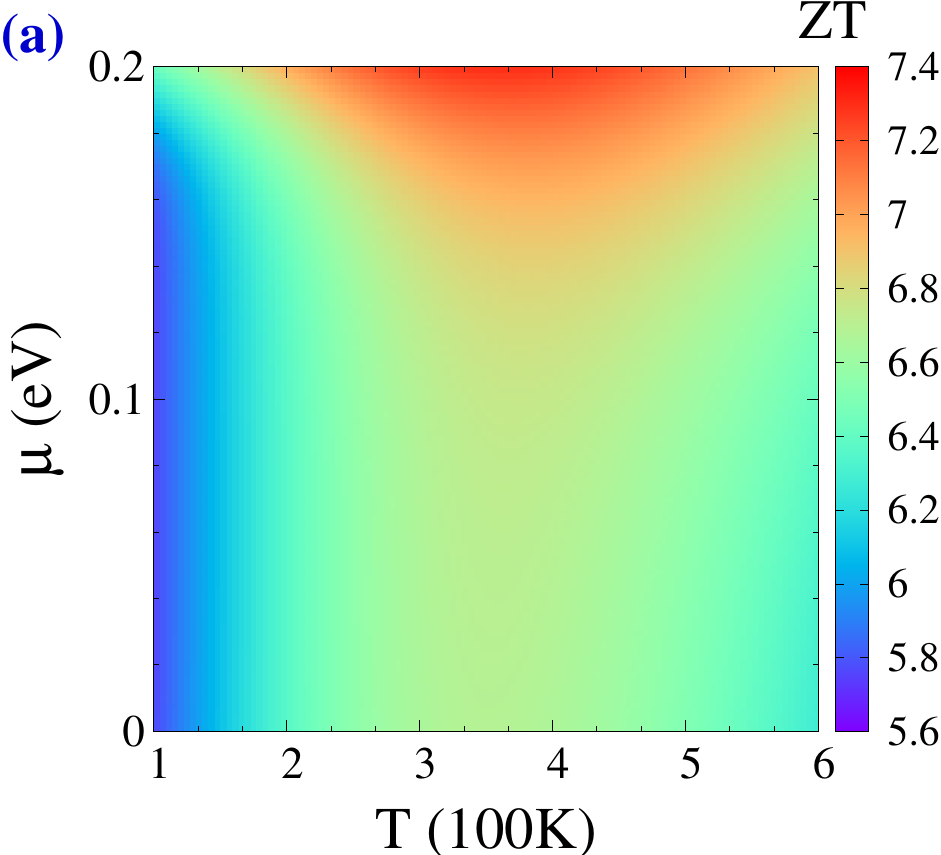}\hspace{0.15cm}\includegraphics[height=4.1cm]{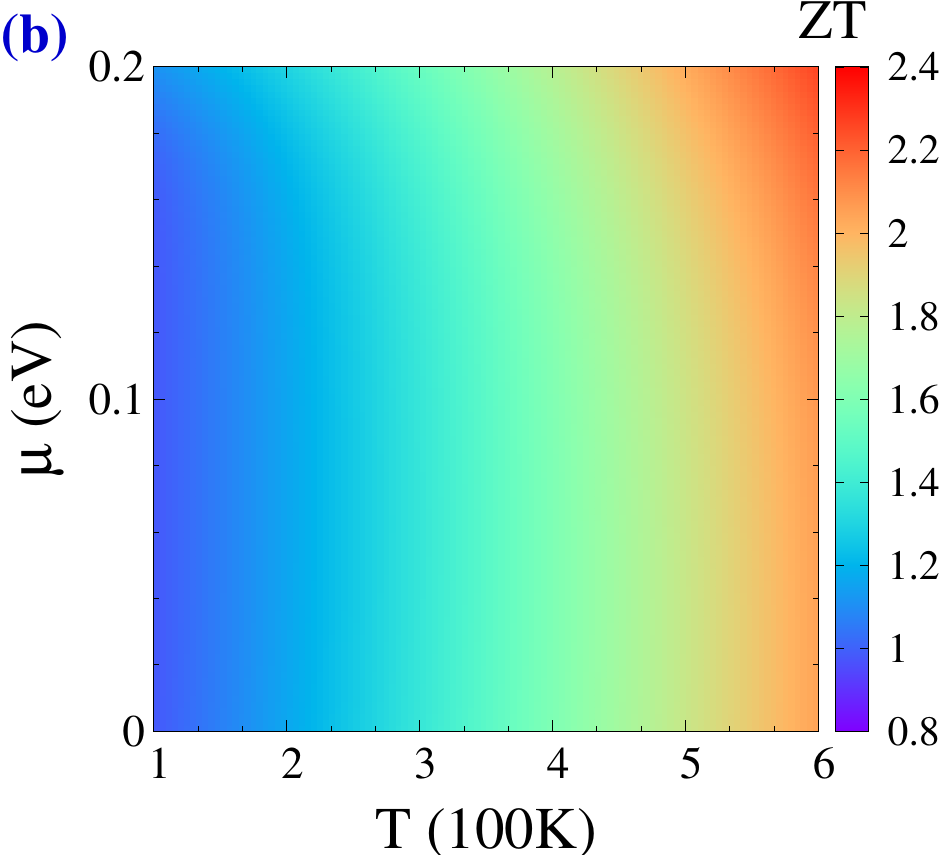}}
  \centering \hspace{-0.44cm}  \includegraphics[width=4.2cm]{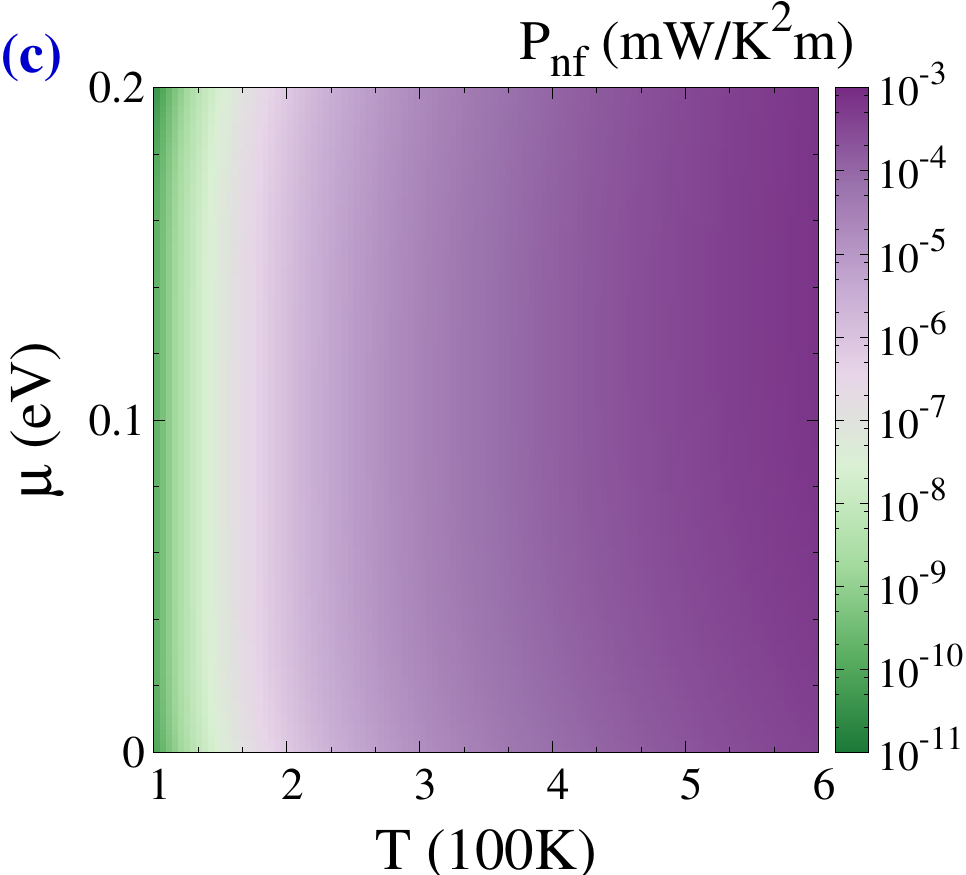}\hspace{0.25cm}\includegraphics[height=3.7cm]{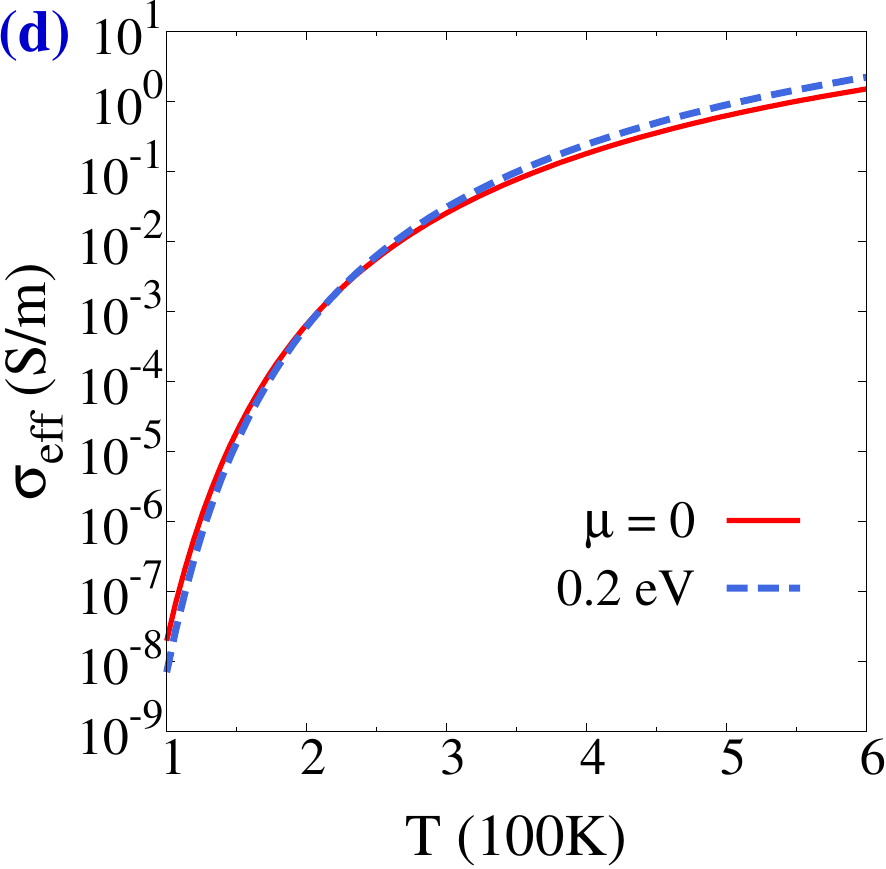}
  \caption{ (Color online) {Thermoelectric figure-of-merit, power-factor, and electrical conductivity of the NFI3T heat engine
  using InSb as the absorption material. (a) and (b), thermoelectric figure-of-merit for the parameters $\Lambda_{nf}/E_g^2=0.2$ 
  and 1.2, respectively. (c) The power-factor $P_{nf}$ and (d) the electrical conductivity 
  $\sigma_{eff}$ of the NFI3T heat engine. The vacuum gap is $d=100$~nm.}}
\label{fig6}
\end{figure}

{While quantifying the parasitic heat conduction needs material and device details, which is rather hard to achieve in theory,
we study the effects of parasitic heat conduction through the parametrization, $K_{para}=G_{ve}\Lambda_{nf}/e^2$, where 
$K_{para}$ is the heat conductance and $\Lambda_{nf}$ is a parameter of the dimension energy square. We calculate the 
figure-of-merit for two cases with parasitic heat conduction parameters $\Lambda_{nf}=0.2E_g^2$ and $\Lambda_{nf}=1.2 E_g^2$. 
Since $E_g$ is much larger than the thermal energy $k_BT$, the latter case corresponds to very strong parasitic heat conduction. }

{We find that for the case with smaller parasitic heat conduction the  optimal figure-of-merit is reached around 350~K 
for electron-doped InSb [see Fig.~6(a)]. The optimal figure-of-merit can be as large as $ZT>7$. For the case with 
$\Lambda_{nf}=1.2 E_g^2$ the optimal figure-of-merit is reached around 600~K for electron-doped InSb with a large value of $ZT>2$
within the range of calculation [see Fig.~6(b)]. From these results, one can see that the optimization of thermoelectric performance 
strongly depends on the parasitic heat conduction which needs to be examined in the future. Figs.~6(c) and 6(d) give the 
dependences of the power factor $P_{nf}=\sigma_{eff}S^2$ and the charge conductivity 
$\sigma_{eff}$ (determined from the total conductance $G_{eff}$ and the cross-section ${\cal A}$ and the thickness $l$ of the device, 
via $\sigma_{eff}=lG_{eff}/{\cal A}$) on the temperature and chemical potential, respectively. Both of them increase rapidly with the 
temperature, because the thermal  distributions of the infrared photons with $\hbar\ome>E_g$ are
significantly increased at elevated temperatures. At high temperatures, the power factor and the charge conductivity can be considerable. 
Roughly, a good balance of the
figure-of-merit and the power-factor can be found in the region of $E_g/7k_B<T<E_g/2k_B$.
In Appendix B, we include the temperature-dependance of the band-gap of InSb and study the temperature and doping dependences of
thermoelectric figure-of-merit. }

\section{Conclusion and Outlook}
We propose a powerful and efficient inelastic thermoelectric QDs heat engine based on near-field enhanced radiative heat transfer. 
By introducing two continua (separated by a spectral gap $E_g$) which are connected to the left and right QDs layers, respectively, 
we introduce a ratchet mechanism through photo-carriers generation and conduction in the infrared regime. The only way that heat 
leaves the hot bath is by exciting electrons from the lower to the upper continuum. The injected thermal radiation then 
induces a directional charge flow and generates electrical power, manifesting as a three-terminal thermoelectric effect. Introducing the 
two continua and the near-field radiative heat transfer substantially increases the inelastic transition rates and hence the
output power. Compared with previous designs of direct photon-assisted hopping between QDs, the thermoelectric heat engine
proposed here is much more powerful. Using a narrow band gap semiconductor InSb as the absorption material, we show that the 
photo-current and the output power can be considerably large. The Seebeck coefficient and figure-of-merit for the proposed heat engine
is significantly large, even when parasitic heat conduction (i.e., conduction of heat through unused photons) is taken into account.
Near-field effects can improve the absorbed heat flux by nearly two orders-of-magnitude, leading to strong enhancement of output
power.

{Roughly, around the regime with $E_g/6k_B<T<E_g/2k_B$ gives a good balance of thermoelectric figure-of-merit and power factor.
Specific optimization depends rather on the transport details and material properties of the heat engine. However, there are
several guiding principles in the search of promising materials for near-field inelastic thermoelectric heat engine. Beside a suitable band gap, 
the inter-band transitions must be efficient which can be realized by a large dipole matrix element or a large joint density
of states for the lower and upper bands. The near field heat transfer must be optimized by utilizing advanced optical structures and
materials, such as hyperbolic metamaterials. Using optical means, near-field heat radiation can be controlled more effectively, compared
to heat diffusion, which provides opportunities for high-performance thermoelectric devices. Although the need 
for a submicron vacuum gap is technologically challenging, recent studies have shown that near-field radiative heat transfer can
also be realized {\em without} a vacuum gap \cite{nfr1,nfr2}, opening opportunities for future thermoelectric technologies based on
inelastic transport mechanisms. Our study presents analog and comparison between photon-induced inelastic thermoelectricity to 
conventional thermoelectricity, which already gives promising results in the linear-response regime and will serve as the foundations for future studies.}

\section*{Acknowledgment}
JHJ acknowledges supports from the National Natural Science Foundation of China
(no. 11675116) and the Soochow University. He also thanks Weizmann
Institute of Science for hospitality and Ming-Hui Lu for helpful discussions. YI acknowledges supports from the US-Israel Binational 
Science Foundation (BSF) and the Weizmann Institute of Science.

\appendix

\section{Computing Interband Transition Rates}
The transition rate in Eq.~(\ref{gam0}) is the central quantity of interest. We show below how it can be calculated numerically.
Consider a thin film of the absorbing material with area ${\cal A}$ and thickness $l_{ab}$, then
\begin{align}
 \Gamma_0(\ome) =& 2\pi \nu_{ph}(\ome) {\cal F}_{nf} (\ome) \sum_{{\vec q}} |g(\ome)|^2
\delta(E_{e,{\vec q}}-E_{v,{\vec q}}-\hbar\ome) \nn\\
&\times f^0(E_{v,{\vec q}}, T)[1-f^0(E_{e,{\vec q}}, T)]N^0(\ome, T) 
\end{align}
For parabolic bands with dispersions given in Eq.~(\ref{insb2}), energy conservation gives,
\begin{align}
E_{e,q} = E_g + \frac{m_{ev}(\hbar\ome-E_g)}{m_e}, \quad E_{v,q} = - \frac{m_{ev}(\hbar\ome-E_g)}{m_v} ,\nn
\end{align}
where $m_{ev}=(m_e^{-1}+m_v^{-1})^{-1}$. The integral over $q$ can be carried out analytically,
\begin{align}
 \Gamma_0(\ome) =& \frac{m_{ev}}{\pi\hbar^3} \nu_{ph}(\ome) {\cal F}_{nf} (\ome) |g(\ome)|^2
{\cal A}l_{ab}  \sqrt{2m_{ev}(\hbar\ome-E_g)} \nn\\
&\times f^0(E_{v, q})[1-f^0(E_{e,q})]N^0(\ome, T) .
\end{align}
The integral over $\ome$ can be carried out numerically from $E_g/\hbar$ to a sufficiently large energy cut-off.

\begin{figure}[htb]
  \centerline{ \includegraphics[width=4.2cm]{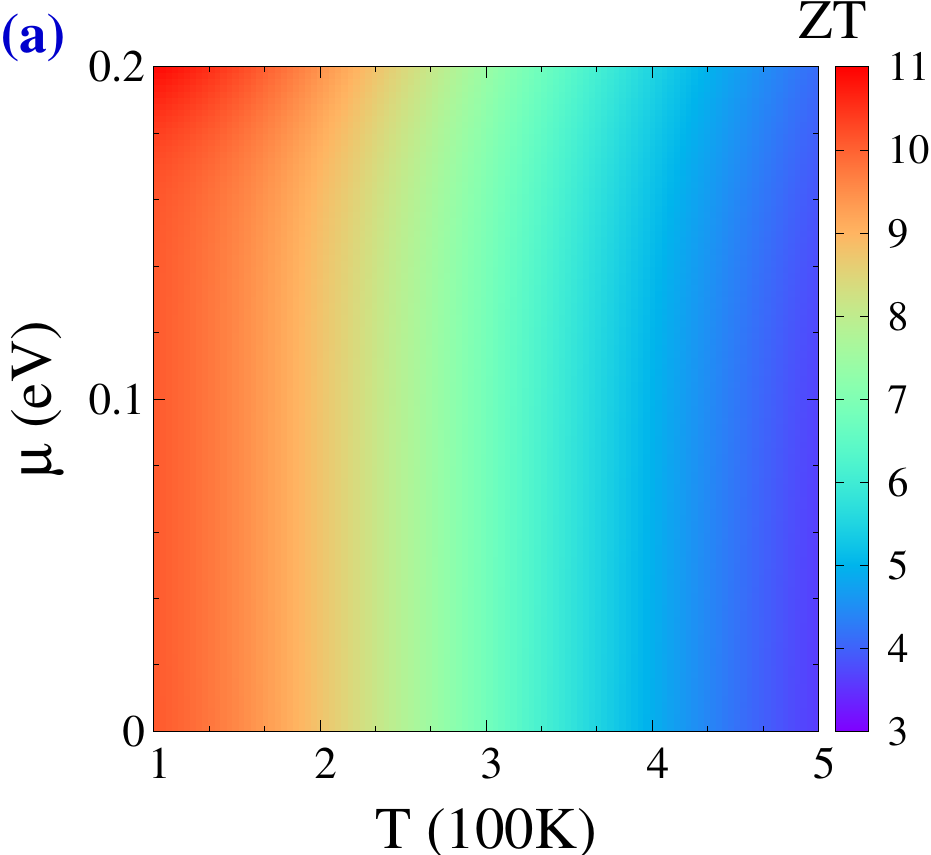}\hspace{0.2cm}\includegraphics[width=4.2cm]{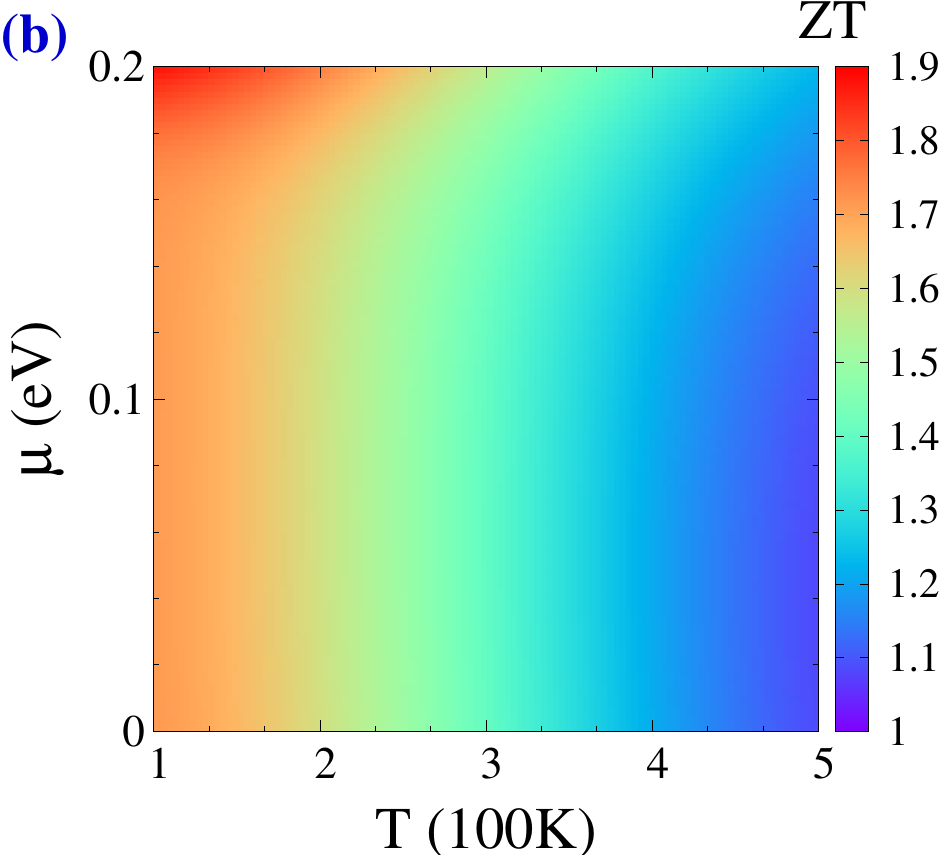}}
  \centerline{ \includegraphics[width=4.2cm]{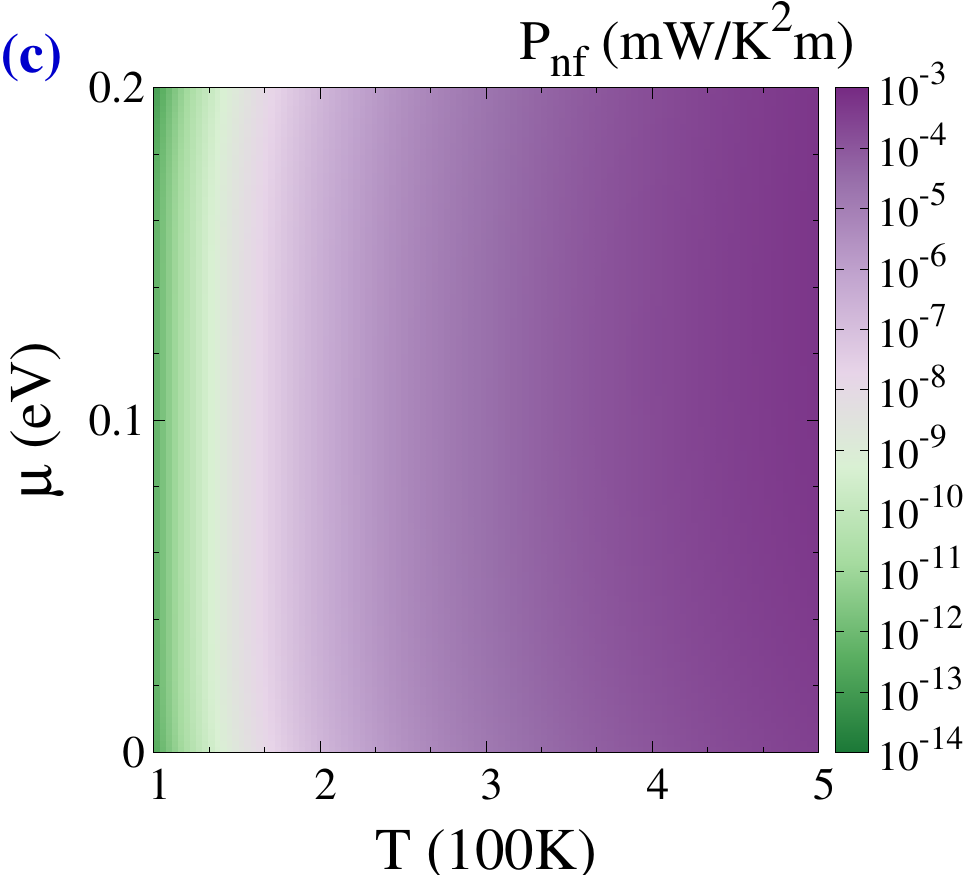}\hspace{0.2cm}\includegraphics[width=4.2cm]{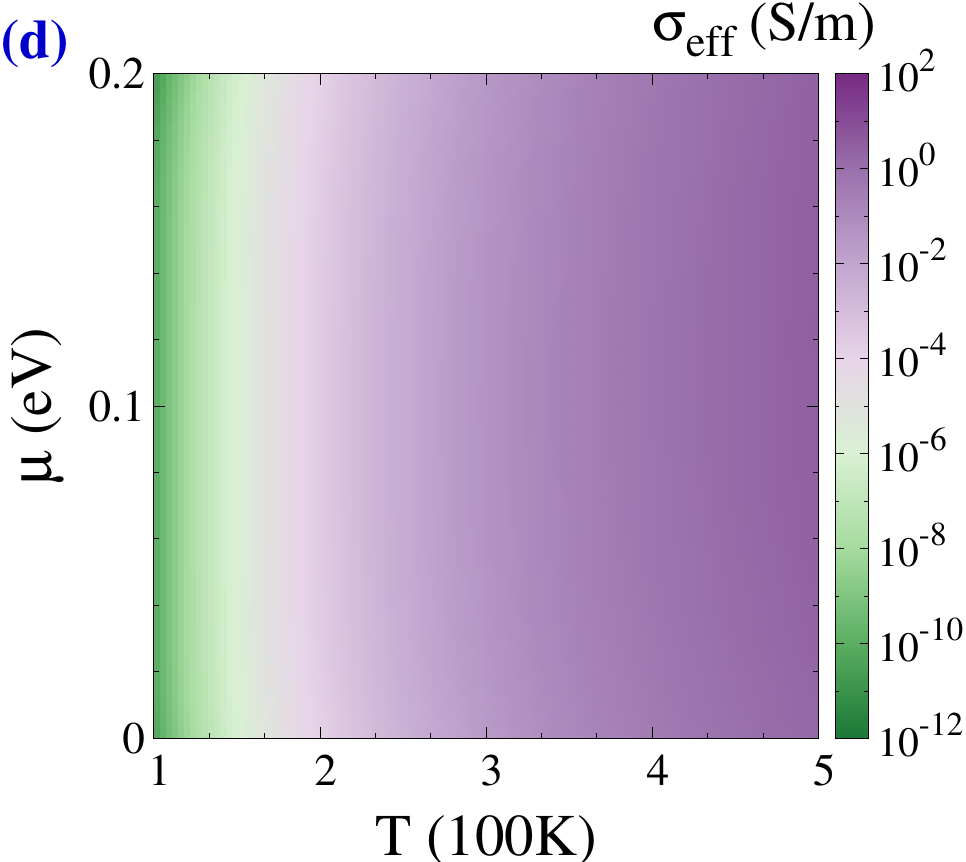}}
   \caption{(Color online) {(a) and (b), thermoelectric figure-of-merit as functions of the chemical potential and temperature, 
   for the parameters $\Lambda_{nf}/E_g^2=0.2$ and 1.2, respectively. (c) and (d) thermoelectric power factor $P_{nf}$ and charge conductivity 
   $\sigma_{eff}$ as functions chemical potential and temperature, respectively. The vacuum gap is $d=100$~nm. The temperature dependance 
   of the band gap $E_g$ of InSb is taken into account.}}
\label{fig7}
\end{figure}

\section{Effects of temperature-dependance of the InSb band gap}
{Here we study the effects of temperature dependent band gap of the absorption material, InSb, on thermoelectric figure-of-merit.
The temperature dependance of the band gap is given by the empirical law of $E_g= 0.24 - 6\times 10^{-4}T^2/(T+500)$ as 
found in Ref.~\onlinecite{norm} for the temperature range $0<T<300$~K. Here $E_g$ is in unit of eV, while the temperature $T$ is in unit of 
Kelvin. We assume this temperature dependance is approximately applicable for the range of interest $100<T<500$~K.
With such temperature-dependance taken into account, we recalculate the temperature and chemical-potential dependances 
of thermoelectric figure of merit. The results for $\Lambda_{nf}=0.2E_g^2$ and $1.2E_g^2$ are presented in Figs.~7(a) and 7(b), respectively.
We find that the optimal conditions for the figure of merit are modified for both cases. Besides, the optimal figure of merit for  $\Lambda_{nf}=0.2E_g^2$
is increased to 11, whereas the optimal figure of merit for $\Lambda_{nf}=1.2E_g^2$ is reduced to 1.9. The power-factor and the charge
conductivity is optimal in the high temperature regime, whereas the figure-of-merit is optimal in the low-temperature regime. A 
balanced optimization of both the figure-of-merit and the power-factor can be found in the region of $E_g/6k_B<T<E_g/2k_B$.
Our calculation here demonstrates again that optimization of thermoelectric performance depends on the transport details of the heat engine.}

{}

\end{document}